\documentclass[a4paper,fleqn,usenatbib,useAMS]{mnras}

\usepackage{graphicx}	
\usepackage{amsmath}	
\usepackage{amssymb}	
\usepackage{multicol}        
\usepackage{bm}		
\usepackage{pdflscape}	
\usepackage{multirow}
\usepackage{xcolor}
\usepackage{hyperref}
\usepackage{booktabs}
\usepackage{subfigure}

\def\specialname[#1]{\textbf{\textsc{#1}}}

\usepackage[T1]{fontenc}
\usepackage{ae,aecompl}

\usepackage{newtxtext,newtxmath}

\title[Relating galaxies across different redshfit]{
Relating galaxies across different redshift to study galaxy evolution}
\author[Wang et al.]{
	Kai Wang,$^{1,2}$\thanks{Contact e-mail: wkcosmology@gmail.com\href{}{}}
    \thanks{Present address: Kavli Institute for Astronomy and Astrophysics, Peking University, Beijing 100871, China}
    Houjun Mo,$^{3}$ Cheng Li$^{2}$ and Yangyao Chen$^{4, 5}$
	\\
	$^{1}$Kavli Institute for Astronomy and Astrophysics, Peking University, Beijing 100871, China\\
	$^{2}$Department of Astronomy, Tsinghua University, Beijing 100084, China\\
	$^{3}$Department of Astronomy, University of Massachusetts Amherst, MA 01003, USA\\
	$^{4}$Key Laboratory for Research in Galaxies and Cosmology, Department of Astronomy, University of Science and Technology of China, Hefei, Anhui 230026, China\\
	$^{5}$School of Astronomy and Space Science, University of Science and Technology of China, Hefei 230026, China
}

\date{Last updated 2022 October 25; in original form 2022 October 25}

\pubyear{2022}

\begin{document}
	\label{firstpage}
	\pagerange{\pageref{firstpage}--\pageref{lastpage}}
	\maketitle


\begin{abstract}
    We propose a general framework leveraging the galaxy-halo connection to
    link galaxies observed at different redshift in a statistical way, and use
    the link to infer the redshift evolution of the galaxy population. Our
    tests based on hydrodynamic simulations show that our method can accurately
    recover the stellar mass assembly histories up to $z\sim 3$ for present
    star-forming and quiescent galaxies down to $10^{10}h^{-1} M_{\odot}$.
    Applying the method to observational data shows that the stellar mass
    evolution of the main progenitors of galaxies depends strongly on the
    properties of descendants, such as stellar mass, halo mass, and star
    formation states. Galaxies hosted by low-mass groups/halos at the present
    time have since $z\sim 1.8$ grown their stellar mass $\sim 2.5$ times as
    fast as those hosted by massive clusters. This dependence on host halo mass
    becomes much weaker for descendant galaxies with similar star formation
    states. Star-forming galaxies grow about 2-4 times faster than their
    quiescent counterparts since $z\sim 1.8$. Both TNG and EAGLE simulations
    over-predict the progenitor stellar mass at $z>1$, particularly for
    low-mass descendants.
\end{abstract}




\section{Introduction}%
\label{sec:introduction}

One goal in galaxy formation and evolution is to establish a predictive
formulation of the assembly histories of galaxies and their dark matter halos
so as to understand the evolution of galaxies in their stellar populations and
gas components
\citep{nationalacademiesofsciencesPathwaysDiscoveryAstronomy2021}. One way to
do this is to establish connections for galaxies observed at different
redshift. Galaxy surveys allow us to study galaxy populations in our Universe
at different cosmic times \citep{2000AJ....120.1579Y,
    lillyZCOSMOS10kBRIGHTSPECTROSCOPIC2009, gerkeDEEP2GalaxyRedshift2007a,
    driverGalaxyMassAssembly2022, tomczakGALAXYSTELLARMASS2014,
takadaExtragalacticScienceCosmology2014, maiolinoMOONRISEMainMOONS2020}.
However, galaxies at different redshift are not causally connected (see blue
points on the left side of Fig.\,\ref{fig:figure/summary}). Mathematically,
galaxy surveys only provide us with the time-dependent distribution function of
galaxy properties, $f(\mathbf{g}\mid t)$, where $\mathbf{g}$ is the `vector'
representing an array of galaxy properties, such as stellar mass, star
formation rate, luminosity, and color. The evolution of individual galaxies is
then given by $\mathbf{g}(t)$ as a function of $t$. Unfortunately, we can not
obtain $\mathbf{g}(t)$ for individual galaxies observationally. In practice,
the observed galaxy populations at different redshift may be connected in a
statistical sense, and we can obtain constraints on $\mathbf{g}(t)$
statistically.

In the framework provided by the current cosmology, galaxies form and grow in
the gravitational potential of dark matter halos \citep{White1978,
MoBoschWhite2010}. Thus the formation history of a galaxy is expected to be
closely related to the assembly history of its host halo. The assembly of a
halo is characterized by its merger tree, which describes how the progenitors
merge together to form the descendant halo, as illustrated on the right side of
Fig.\,\ref{fig:figure/summary}. The formation of the halo population in the
current cosmological framework is well understood, and large samples of halo
merger trees are now available from semi-analytical models and cosmological
simulations of dark matter halos. Clearly, the information provided by halo
merger trees should be used as a key component in modeling the assembly of
galaxies.

In order to use halo merger trees to follow galaxy assembly, it is necessary to
link the observed galaxy populations with the halo population. Three approaches
have been used to make the link. The first one is to simulate the evolution of
different matter components based on first principles, with a set of empirical,
subgrid recipes to model some crucial but unresolved processes
\citep[e.g.][]{vogelsbergerModelCosmologicalSimulations2013,
    schayeEAGLEProjectSimulating2015a,
    vogelsbergerCosmologicalSimulationsGalaxy2020,
nelsonIllustrisTNGSimulationsPublic2019}. The second approach is to model
galaxy evolution processes through a set of parametric functions, with the
model parameters constrained using summary statistics obtained from galaxy
surveys, such as the distribution functions of galaxies with respect to
intrinsic properties \citep[e.g.][]{liDistributionStellarMass2009,
    baldryGalaxyMassAssembly2012, moustakasPRIMUSCONSTRAINTSSTAR2013,
davidzonCOSMOS2015GalaxyStellar2017, chenELUCIDVICosmic2019} and spatial
correlation functions of galaxies
\citep[e.g.][]{liDependenceClusteringGalaxy2006}. This approach can be further
divided into two sub-categories: semi-analytical models
\citep[e.g.][]{White1991, 1999MNRAS.310.1087S,
    kangSemianalyticalModelGalaxy2005, Croton2006,
    deluciaHierarchicalFormationBrightest2007, bensonGalaxyFormationTheory2010,
guoDwarfSpheroidalsCD2011, henriquesGalaxyFormationPlanck2015} and empirical
models \citep{zhengGalaxyEvolutionHalo2007, 2007ApJ...655L..69W,
    conroyCONNECTINGGALAXIESHALOS2009,
    mosterConstraintsRelationshipStellar2010, yangEVOLUTIONGALAXYDARK2012,
    luEmpiricalModelStar2014, behrooziUniverseMachineCorrelationGalaxy2019,
kipperRoleStochasticSmooth2021}, depending on whether the functions
describing galaxy evolution processes are physically or empirically
motivated. The third approach is to establish the galaxy-halo connection
through some heuristic assumptions.  For instance,
\citet{vandokkumGROWTHMASSIVEGALAXIES2010} derived the stellar mass
evolution histories for massive galaxies assuming that the stellar mass
rank of a galaxy is preserved during the time period in question
\citep{brownEvolvingLuminosityFunction2007,
    jaacksCONNECTINGDOTSTRACKING2016,
talOBSERVATIONSENVIRONMENTALQUENCHING2014}. This assumption is expected to
be invalid if galaxy mergers are important during the period of interest,
or if the variance in the assembly histories is large for the galaxy
population in concern \citep{lejaTracingGalaxiesCosmic2013}.
\citet{behrooziUSINGCUMULATIVENUMBER2013} refined this method and proposed
the use of an evolving cumulative number density based on the sub-halo
abundance matching method \citep[see
also][]{moStructureClusteringLymanbreak1999,
    yangConstrainingGalaxyFormation2003, Vale2004,
    conroyModelingLuminosityDependent2006, guoHowGalaxiesPopulate2010,
    behrooziCOMPREHENSIVEANALYSISUNCERTAINTIES2010, liStellarMassStellar2013,
2021MNRAS.508..175C}. In this method, the evolution in the mass rank of
subhalos obtained from $N$-body simulations is used to mimic that in the
stellar mass rank of galaxies \citep{torreyAnalysisEvolvingComoving2015,
    torreyForwardBackwardGalaxy2017, wellonsImprovedProbabilisticApproach2017,
hillMassColorStructural2017}, and it has been shown that this method is
able to recover the stellar mass evolution of the main progenitors of
galaxies from $z\gtrsim 3$ to $z\sim 0$ in the EAGLE and Illustris
simulations \citep{clauwensLargeDifferenceProgenitor2016,
torreyAnalysisEvolvingComoving2015}. The methods based on the cumulative
number density of galaxies have been applied to infer the stellar mass
growth of massive galaxies and the evolution of their star formation
activities and structures \citep{vandokkumGROWTHMASSIVEGALAXIES2010,
hillMassColorStructural2017}. However, as found by
\citet{clauwensLargeDifferenceProgenitor2016}, the stellar mass evolution
depends significantly on the star formation states of descendant galaxies
at $z=0$ in the EAGLE simulation, in the sense that passive descendants
have more massive progenitor galaxies, especially for low-mass galaxies,
and the cumulative number density method is unable to reproduce the trend.

In this paper, we leverage the established galaxy-halo connection to propose a
framework to connect galaxies across cosmic time. The most basic relation
between galaxies and dark matter halos is the central galaxy stellar mass- halo
mass relation (SMHM relation), which may be described by some parametric forms
\citep{yangGALAXYGROUPSSDSS2009, guoHowGalaxiesPopulate2010,
    behrooziCOMPREHENSIVEANALYSISUNCERTAINTIES2010,yangEVOLUTIONGALAXYDARK2012,
wechslerConnectionGalaxiesTheir2018}. With more information gathered from
observations and numerical simulations, it is also possible to study secondary
effects in the  galaxy-halo connection, and models making use of galaxy and
halo properties in addition to mass have been proposed
\citep[e.g.][]{hearinDarkSideGalaxy2013, hearinDarkSideGalaxy2014,
watsonPredictingGalaxyStar2015, behrooziUniverseMachineCorrelationGalaxy2019}.
Although a consensus is still to be reached, some promising results have been
obtained from such modeling. For instance, relating galaxy color to halo
formation time as a secondary constraint on the galaxy-halo connection,
\citet{hearinDarkSideGalaxy2013} were able to reproduce the clustering
properties for galaxies and their dependence on galaxy color and star formation
activity \citep[see also][]{watsonPredictingGalaxyStar2015}. These results will
be used in our method to improve the connection of galaxies across cosmic time
and to alleviate some problems in some earlier models. In particular, we
connect galaxies at different redshift chronologically by populating them into
subhalo merger trees according to the galaxy-halo connection introduced above.
Our incorporation of halo formation times to model galaxy star formation
activities also enables us to follow galaxy assembly histories conditioned on
the star formation state of descendant galaxies, thus differentiating between
active and passive descendants.

The paper is organized as follows. In \S\,\ref{sec:method}, we introduce
our framework. We test our method using hydrodynamic simulations in
\S\,\ref{sec:performance}. We apply our method to observational data in
\S\,\ref{sec:application}. Finally, we summarize our results in
\S\,\ref{sec:summary}.

\section{Method to link galaxies with their progenitors}%
\label{sec:method}

\subsection{The framework of connecting galaxies across cosmic time}%
\label{sub:framework}

\begin{figure*}
    \centering
    \includegraphics[width=0.8\linewidth]{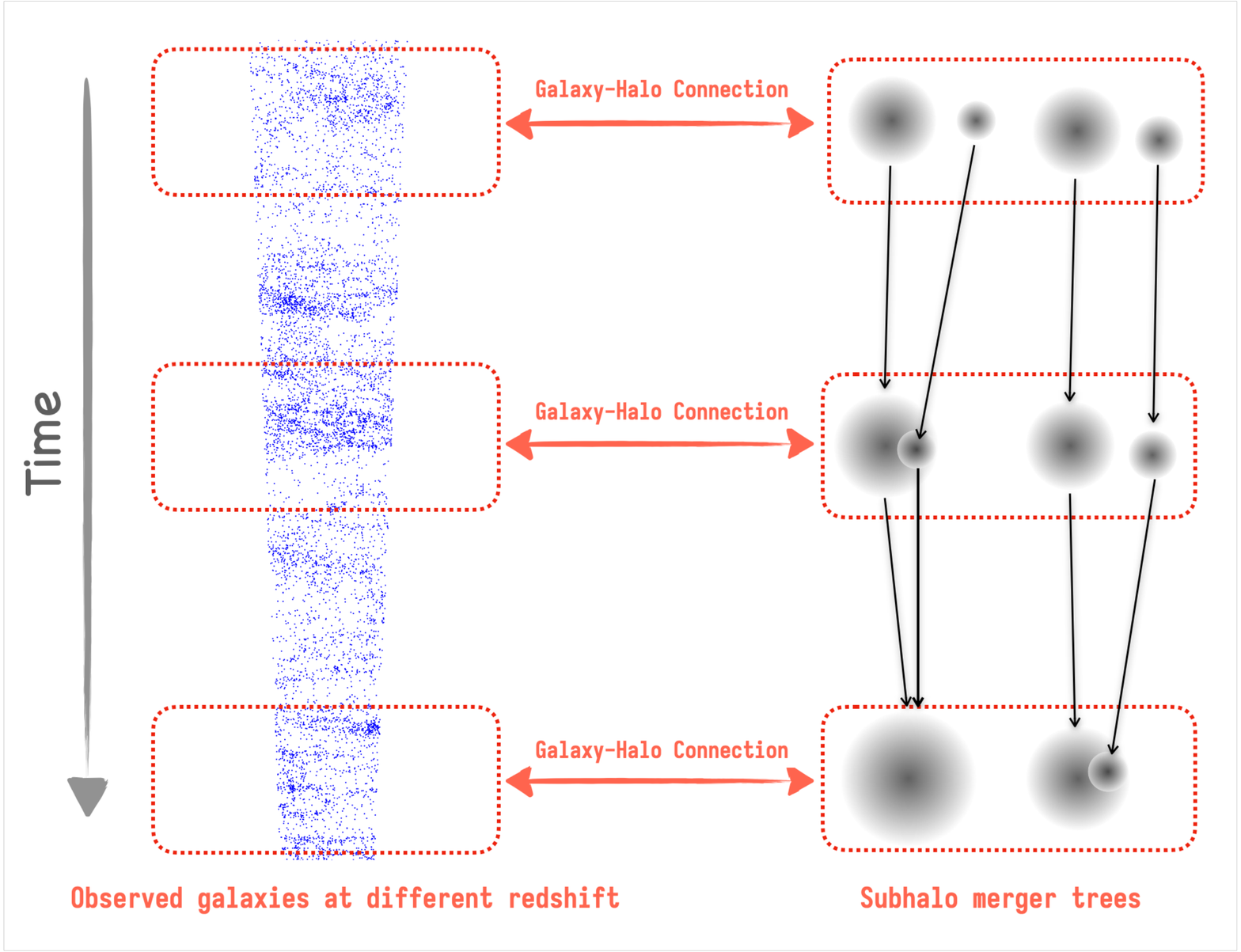}
    \caption{
        Demonstration of the method that connects galaxies across cosmic time.
        On the left hand, blue points are the observed galaxies from high
        redshift (top) to low redshift (bottom). On the right hand, black
        shades are halos and subhalos on subhalo merger trees from N-body
        simulations. We first connect galaxies with halos using established
        galaxy-halo connections, then use links in subhalo merger trees to
        connect galaxies across cosmic time.
    }%
    \label{fig:figure/summary}
\end{figure*}

 Here we propose a general framework to connect galaxies across cosmic time. In
 Fig.\ \ref{fig:figure/summary}, the left side shows observed galaxies at
 different redshift, from which we can calculate the distribution of galaxy
 properties as a function of redshift. The right side shows subhalo merger
 trees constructed from N-body simulations or with semi-analytical methods,
 where each subhalo can have multiple progenitors but only one descendant.
 These connections form the so-called subhalo merger trees. In the current
 scenario of galaxy formation, galaxies form and evolve in dark matter
 halos/subhalos. It is thus possible to incorporate the properties of subhalo
 merger trees when linking galaxies at different cosmic times. This can be done
 as follows. First, we use the relation between the subhalo and galaxy
 populations at a given redshift to link galaxies to subhalos at that redshift,
 as indicated schematically by the horizontal lines in
 Fig.\,\ref{fig:figure/summary}. Second, we use subhalo merger trees to connect
 galaxies at different redshift according to the subhalos that the galaxies are
 linked to. Finally, we infer the evolution of galaxies based on the link
 established for galaxies at different redshift.

The key piece of this method is the relation between subhalos and galaxy
populations, and this is the main difference between our work and previous
studies \citep{vandokkumGROWTHMASSIVEGALAXIES2010,
behrooziUSINGCUMULATIVENUMBER2013}. \citet{behrooziUSINGCUMULATIVENUMBER2013}
used the SMHM relation to connect galaxies of a given stellar mass rank to
subhalos of the same subhalo mass rank, and to study the stellar mass assembly
histories of galaxies. Here we take a step further by adding the star formation
states of galaxies at $z\sim 0$ into the galaxy-subhalo connection. We can thus
study the stellar mass assembly histories conditioned on the star formation
activities of descendant galaxies. In addition, we highlight the dependence on
the host halo mass of descendant galaxies, which was ignored in previous
studies.

\subsection{Assigning stellar mass to subhalos}%
\label{sub:assigning_stellar_mass_to_subhalos}

One important part of the framework described above is to link galaxies to dark
matter halos at a given redshift. The exact link depends on the details of
galaxy formation and evolution in dark matter halos. For an empirical model
such as the one considered here, the assumptions adopted in establishing this
link need to be sufficiently flexible to allow uncertainties in this link. We
adopt an abundance matching method to link galaxies to subhalos. In this
subsection, we describe our method to link the stellar mass of galaxies to the
mass of subhalos, taking into account the scatter in their relation. In the
following subsection (\S\,\ref{sub:assigning_star_forming_state}), we describe
how to assign star formation states to subhalos.

Sub-halo abundance matching (SHAM) is an empirical method to populate galaxies
into subhalos, assuming a monotonic relation between one of the galaxy
properties (e.g. luminosity and stellar mass) and one of the subhalo properties
(e.g. peak halo mass and peak circular velocity)
\citep{moStructureClusteringLymanbreak1999,
    yangConstrainingGalaxyFormation2003, Vale2004,
    conroyModelingLuminosityDependent2006, guoHowGalaxiesPopulate2010,
    behrooziCOMPREHENSIVEANALYSISUNCERTAINTIES2010, liStellarMassStellar2013,
2021MNRAS.508..175C}. Here we use the stellar mass of galaxies, $M_*$, and the
peak halo mass of subhalos\footnote{The peak halo mass of a subhalo is defined
    as the maximum halo mass that a subhalo has achieved along its main branch
when it is a central subhalo.}, $M_{\rm p}$, to perform the SHAM procedure.
As found in previous studies, it is important to take into account the
variance in the relation between galaxy and halo properties in order to
better reproduce summary statistics of the observed galaxy population, such
as the two-point correlation function
\citep{behrooziCOMPREHENSIVEANALYSISUNCERTAINTIES2010,
reddickCONNECTIONGALAXIESDARK2013, hearinSHAMClusteringNew2013}. Our model
includes such variances. The method consists of the following steps:
\begin{enumerate}

    \item Perform the SHAM procedure for galaxies at $z=0$ using
        \begin{equation}
            n(>M_{*, 0}) = n(>M_{\rm p, 0})\,, \label{eq:am_0}
        \end{equation}
        where the left-hand side is the cumulative stellar mass function of
        galaxies at $z=0$, and the right-hand side is the cumulative peak mass
        function of subhalos at $z=0$. This matching assigns a \emph{tentative}
        stellar mass, $M_{\rm d, 0}$, to a subhalo according to its peak mass.
        To take into account the scatter in the SMHM relation, we define for
        each subhalo a mass indicator, $M_{\rm ind, 0}$:
        \begin{equation}
            \log M_{\rm ind, 0} = \log M_{\rm d, 0} + \epsilon_0\,,
            \label{eq:tentativa_0}
        \end{equation}
        where $\epsilon_0$ is a random number generated from some distribution
        function $P_0(\epsilon_0)$. In our model, we assume $P_0$ to be a
        normal distribution with zero mean and a dispersion of $\sigma_0=0.2$,
        as motivated by the SMHM relation inferred from observation
        \citep[e.g.][]{yangGALAXYGROUPSSDSS2009,
            behrooziCOMPREHENSIVEANALYSISUNCERTAINTIES2010,  Li2012,
            reddickCONNECTIONGALAXIESDARK2013,
        wechslerConnectionGalaxiesTheir2018}. This mass is used to re-rank all
        subhalos. The stellar mass of a galaxy is then assigned to a subhalo of
        the same rank using
        \begin{equation}
            n(>M_{*, 0}) = n(>M_{\rm ind, 0})\,.
        \end{equation}
        The assigned stellar mass so obtained is denoted as $M_{\rm sca, 0}$,
        where `sca' stands for `scatter'. Note that $M_{\rm sca, 0}$ has the
        same distribution as $M_{\rm d, 0}$, but is no longer a monotonic
        function of $M_{\rm p, 0}$ because of the nonzero scatter.

    \item Perform the same SHAM procedure for galaxies at high $z$ using
        \begin{equation}
            n(>M_{*, z}) = n(>M_{{\rm p}, z})\,,
        \end{equation}
        and we denote the assigned {\it tentative} stellar mass as $M_{{\rm d},
        z}$. To add scatter in the SMHM relation at high $z$, we again define a
        mass indicator, $M_{{\rm ind}, z}$:
        \begin{equation}
            \log M_{{\rm ind},z} = \log M_{{\rm d},z} + \epsilon_z\,,
            \label{eq:tentative}
        \end{equation}
        with $\epsilon_z$ being a random variable specified by a probability
        distribution function $P_z(\epsilon_z)$, and use it to re-rank
        subhalos. Since we want to link a $z=0$ galaxy with a given stellar
        mass and subhalo mass to its progenitors at higher $z$,
        $P_z(\epsilon_z)$ is conditional on the descendant. For simplicity, we
        model $P_z$ with a normal distribution, and describe the correlation
        with the descendant through a correlation coefficient $\rho_z$. Thus,
        we can write
        \begin{align}
            &P(\epsilon_z\mid M_{\rm sca, 0},~M_{\rm d, 0})\nonumber\\
            &\quad= \mathcal{N}\left(\frac{\rho_z \sigma_z\log(M_{\rm sca, 0} /
            M_{\rm d, 0})}{\sigma_0},~(1 - \rho_z^2)\sigma_z^2\right)\,,
        \end{align}
        where $\mathcal{N}(\mu,~\sigma)$ is a normal distribution with mean
        $\mu$ and standard deviation $\sigma$, and $\rho_z$ describes the
        correlation between $\log(M_{{\rm ind}, z} / M_{{\rm d}, z})$ and $\log
        (M_{\rm sca, 0}/M_{\rm d, 0})$, i.e. the correlation of the deviation
        of the stellar mass from the tentative mass at $z$ with that of its
        descendant at $z=0$. We use results obtained from the TNG simulation to
        calibrate the model (see Appendix
        \ref{sec:scatter_of_stellar_mass_halo_mass_relation_in_tng300_1}). As
        shown in the left panel of
        Fig.\,\ref{fig:figure/sm_hm_relation_in_TNG}, the scatter in the SMHM
        relation, $\sigma_z$, is independent of redshift, so that
        $\sigma_z\approx \sigma_0$ and
        \begin{equation}
            P(\epsilon_z\mid M_{\rm sca, 0},~M_{\rm d, 0}) =
            \mathcal{N}\left(\rho_z \log(M_{\rm sca, 0} / M_{\rm d, 0}),~(1 -
            \rho_z^2)\sigma_0^2\right)\,.
        \end{equation}
        Thus, $\rho_z=0$ implies that $\epsilon_z$ has a Gaussian distribution
        with zero mean and with dispersion equal to $\sigma_0$, while
        $\rho_z=1$ implies that $\epsilon_z = \log(M_{\rm sca, 0} / M_{\rm d,
        0})$. The right panel of Fig.\,\ref{fig:figure/sm_hm_relation_in_TNG}
        shows that the correlation in the simulation can be described by
        \begin{equation}
            \rho_z = \exp\left(-0.83z\right)\,, \label{eq:rho}
        \end{equation}
        and we adopt it for our model. We note that our results are insensitive
        to $\rho(z)$, as shown in Fig.~\ref{fig:figure/dependence_on_rho_z}.
        Finally, we perform the subhalo abundance matching using
        \begin{equation}
            n(>M_{*, z}) = n(>M_{{\rm ind}, z})\,
        \end{equation}
        to assign the stellar mass to a subhalo, and we denote this mass by
        $M_{{\rm sca}, z}$.

\end{enumerate}

Once subhalos at different redshift are assigned with stellar masses, we link
the stellar mass of a galaxy at $z=0$ to that of its progenitors at higher $z$
using its subhalo merger tree. We can then study the stellar mass distribution
of the progenitors at a given $z$ for galaxies selected at $z=0$.

\subsection{Assigning star-formation states to subhalos}%
\label{sub:assigning_star_forming_state}

Using the EAGLE simulation \citep{schayeEAGLEProjectSimulating2015a},
\citet{clauwensLargeDifferenceProgenitor2016} showed that the progenitor
stellar mass of galaxies selected at $z\sim 0$ depends strongly on the star
formation states of their descendants. Quiescent descendants on average have
more massive progenitors than their star-forming counterparts, and this
dependence is stronger for low-mass descendants. They also found that the
evolving cumulative number density method proposed in
\citet{behrooziUSINGCUMULATIVENUMBER2013} cannot reproduce this trend,
indicating that stellar mass alone is not sufficient to establish a reliable
link between galaxies at different redshift. Some new properties must be
included to improve the method. Here we add the star formation states of
galaxies in the galaxy-halo connection. To this end, we use a conditional age
distribution matching (CADM) method, which assumes a monotonic relation between
the star-forming activity of a galaxy and some characteristic redshift
describing the formation of its subhalo with fixed stellar mass and host halo
mass \citep{hearinDarkSideGalaxy2013, hearinDarkSideGalaxy2014}. We adopt a
characteristic redshift, $z_{\rm starve}$, similar to that used in
\citet{favoleSubhaloAbundanceMatching2022}, where they defined $z_{\rm starve}$
as the maximum among $z_{\rm half}$, $z_{\rm char}$, and $z_{\rm acc}$. Here
$z_{\rm half}$ is the highest redshift when the halo mass of the main
progenitor halo reaches half of the peak halo mass, $M_{\rm p}$; $z_{\rm char}$
is the highest redshift at which the mass of the main progenitor halo reaches
$10^{12}h^{-1}M_{\odot}$; $z_{\rm acc}$ is the redshift at which the subhalo
becomes a satellite subhalo and is set to 0 for all central subhalos. This
definition of $z_{\rm starve}$ is able to reproduce the dependence of galaxy
clustering on star formation rate (SFR) in the TNG simulation, as shown in
\citet{favoleSubhaloAbundanceMatching2022}. Similar definitions have also been
used to interpret the color and SFR dependence of galaxy clustering in
observations \citep[e.g.][]{hearinDarkSideGalaxy2013, hearinDarkSideGalaxy2014,
watsonPredictingGalaxyStar2015}. We found that $z_{\rm half}$ is almost always
greater than $z_{\rm acc}$, and so we use
\begin{equation}
    z_{\rm starve} = \texttt{MAX}\left(z_{\rm char}, z_{\rm half}\right)
\end{equation}
for our analysis.

We proceed to assign star formation states to subhalos at a given redshift
through the following steps:
\begin{enumerate}
    \item Apply the SHAM procedure described in
        \S\,\ref{sub:assigning_stellar_mass_to_subhalos} to assign stellar
        masses to subhalos.
    \item For a fixed stellar mass and host halo mass bin, i.e. $ (M_*\pm
        \Delta M_*/2, M_h\pm \Delta M_h/2)$, we calculate $z_{\rm starve}$ for
        subhalos in this bin, and identify a fraction of $ F_{\rm q}(M_*\pm
        \Delta M_*/2, M_h\pm \Delta M_h/2)$ subhalos with top $z_{\rm starve}$
        as quiescent, while the remaining subhalos are identified as
        star-forming, where $F_{\rm q}$ is the quiescent fraction of galaxies
        in the corresponding stellar mass and host halo mass range.
\end{enumerate}

After assigning stellar masses and star formation states to subhalos at
different redshift, we can use subhalo merger trees to connect them and infer
their redshift evolution. Note that the estimate of $F_{\rm q}$ as a function
of stellar mass and host halo mass requires a complete galaxy group sample with
reliable halo mass assignments. Currently, such samples are available only at
$z\sim 0$.

\section{Testing the method with the TNG simulation}%
\label{sec:performance}

In this section, we use the IllustrisTNG simulation to test the performance
of the method described above. As a check, we also apply our method to the
EAGLE simulation, and the results are presented in
Appendix\,\ref{sec:test_performance_on_eagle_simulation}.

\subsection{Simulation Data}%
\label{sub:data_description}

We use the simulation data from IllustrisTNG (The Next Generation)
\citep[hereafter TNG,][]{nelsonIllustrisTNGSimulationsPublic2019}. TNG is a
suite of gravo-magnetohydrodynamical cosmological simulations run with the
moving-mesh code AREPO.  Compared to the original Illustris, TNG has made a
set of improvements, including an extension of the mass range for the simulated
galaxies and halos, improved numerical and astrophysical modeling, and
addressing identified shortcomings of the Illustris simulation \citep[see][for
the detailed modeling]{pillepichSimulatingGalaxyFormation2018,
weinbergerSimulatingGalaxyFormation2017}. TNG simulates the formation and
evolution of galaxies from $z=127$ to $z=0$ based on a cosmology consistent
with results in \citet{planckcollaborationPlanck2015Results2016}, where
$\Omega_{\Lambda, 0}=0.6911$, $\Omega_{b, 0}=0.3089$, $\sigma_8=0.8159$,
$n_s=0.9667$ and $h=0.6774$. TNG consists of a suite of ten simulations with
different box sizes and resolutions. Here we use TNG100-1, which has $2\times
1820^3$ resolution elements in a box with a side length of $75h^{-1}{\rm cMpc}$.
The target baryon mass resolution is $1.4\times 10^6 M_{\odot}$, and the
dark matter particle mass is $7.5\times 10^6 M_{\odot}$.

In the TNG simulation, dark matter halos were identified with the
friends-of-friends (FoF) algorithm using dark matter particles. Substructures
were identified with the SUBFIND algorithm using all types of particles
\citep{springelPopulatingClusterGalaxies2001}. The baryonic components
identified are considered as galaxies, and the dark matter components as
subhalos. In each FoF halo, the central subhalo is defined as the subhalo
located at the minimum of the gravitational potential, and the galaxy in it is
referred to as the central galaxy. The remaining subhalos are defined as
satellite subhalos, while galaxies in them as satellite galaxies.

Two different merger trees are provided by the TNG simulation: SUBLINK
\citep{rodriguez-gomezMergerRateGalaxies2015a} and LHALOTREE
\citep{springelModellingFeedbackStars2005}. We use SUBLINK trees because of
their easy accessibility. Once all the trees are built up, we can define the
main progenitor of any halo (or galaxy) in the preceding snapshot as the one
with {\it the most massive progenitor history}
\citep[See][]{deluciaHierarchicalFormationBrightest2007}. By recursively
identifying the main progenitor, we obtain a branch consisting of the main
progenitors of the halo in question. This branch is referred to as the main
branch of the halo.

The halo mass we use is the ``top-hat'' mass of the FoF halo within a radius
within which the average overdensity is equal to that predicted by the
spherical collapse model \citep[see][]{1998ApJ...495...80B}. The galaxy stellar
mass we use is defined as the sum of the stellar particles within twice the
stellar half-mass radius, $2R_*$, where $R_*$ is calculated using all the
stellar particles in the sub-halo
\citep{pillepichFirstResultsIllustrisTNG2018}. The galaxy star formation rate
is calculated by summing the birth-time-mass\footnote{The birth-time-mass of
    each stellar particle is different from its current mass due to the mass
loss through stellar wind.} of all the stellar particles formed in the last
200 Myrs within $2R_*$ and divide it by 200 Myrs
\citep{donnariStarFormationActivity2019}. Finally, galaxies are classified
as quiescent when they satisfy
\begin{equation}
    \frac{\rm SFR}{M_*} < 10^{-11}h~{\rm yr}^{-1}, \label{eq:q}
\end{equation}
while the remaining galaxies are classified as star-forming.

To test the performance of our method, we applied it to the TNG simulation to
infer the stellar mass evolution histories and compared them with the results
obtained directly from galaxy merger trees in the simulation. To begin with, we
calculated the stellar mass function at each snapshot and assigned stellar mass
to subhalos using the method described in
\S\,\ref{sub:assigning_stellar_mass_to_subhalos}. We then calculated the
quiescent fraction as a function of the stellar mass and host halo mass of
galaxies at $z=0$ and assigned star formation states to subhalos using the
method described in \S\,\ref{sub:assigning_star_forming_state}. Finally, we
calculated the distribution and redshift evolution of the stellar mass of the
progenitor mass conditioned on the stellar mass and star formation states of
descendant galaxies at $z=0$, and compared the results with those obtained
directly from the TNG simulation. The performances of the method are presented
in relevant subsections in the following.

\subsection{Stellar mass distributions of progenitors}%
\label{sub:progenitor_stellar_mass_distribution}

\begin{figure*}
    \centering
    \includegraphics[width=1\linewidth]{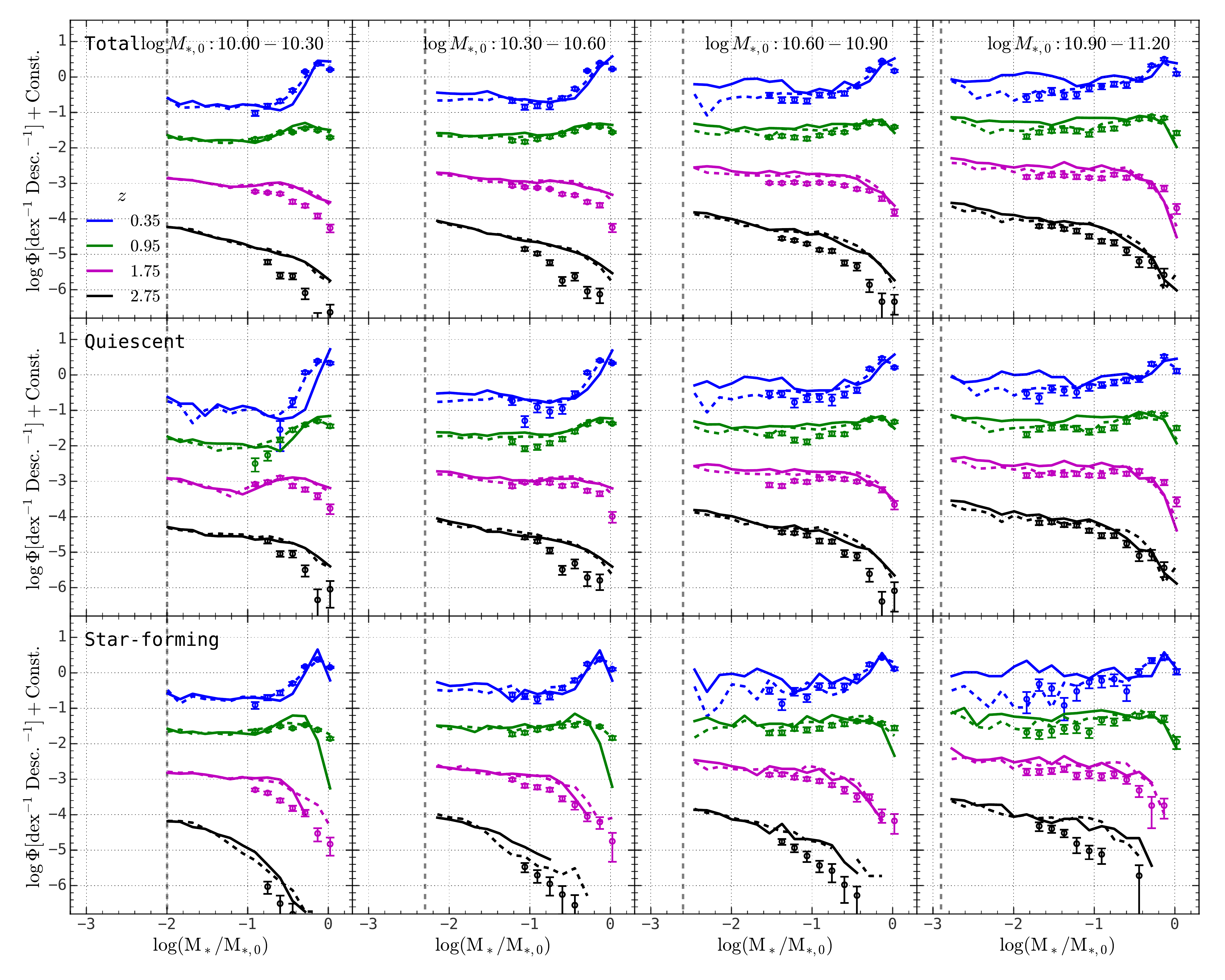}
    \caption{
        Progenitor normalized stellar mass functions for different descendant
        stellar mass bins ({\it different columns}) and different descendant
        star-formation states ({\it different rows}) at four redshift
        snapshots. The solid lines are extracted from galaxy merger trees of
        the TNG simulation, dashed lines are the results of applying our method
        to the TNG simulation. Symbols are the results of applying our method
        to real observation with error bars calculated using the bootstrap
        method. The vertical dashed lines indicate the minimal stellar mass
        resolved by the TNG simulation. Here $y$-values are shifted for clarity.
    }%
    \label{fig:figure/pnsmf_observation_zstarve1}
\end{figure*}

In the current $\Lambda$CDM paradigm, cosmic structures form hierarchically,
and so do galaxies. This hierarchical formation of galaxies can be described
by the distribution of the galaxy population, i.e. the stellar mass function,
at $z=0$ and the distribution of their progenitors at higher $z$, i.e. the
progenitor normalized stellar mass function (PNSMF). The PNSMF is defined as
the number of progenitor galaxies at a given redshift, as a function of the
stellar mass ratio, $M_*/M_{*, 0}$,  for a given set of descendant galaxies
with the same stellar mass $M_{*, 0}$. We first examine the performance of our
method in recovering this statistic, and the result is presented in Fig.\
\ref{fig:figure/pnsmf_observation_zstarve1}, where solid lines are the results
obtained directly from galaxy merger trees in the TNG simulation and dashed
lines are predictions of our method. Results are shown separately for the total
(upper panels), quiescent (middle panels), and star-forming populations (lower
panels). As one can see, the PNSMF is dominated by low-mass galaxies and
declines toward the massive end at high redshift. When evolved to low redshift, the
PNSMF becomes more dominated by massive progenitors. This trend is consistent
with the hierarchical scenario of galaxy assembly where low-mass galaxies
formed at high redshift are assembled into more massive descendants at lower
redshift, and low-mass descendants are assembled earlier than more massive
ones. The redshift evolution of the PNSMF in the simulation is well reproduced
by our method over the entire stellar mass range covered by the TNG simulation,
and for both quiescent and star-forming populations. We also test the
performance of our method on the EAGLE simulation in
Appendix~\ref{sec:test_performance_on_eagle_simulation}, and
Fig.~\ref{fig:figure/pnsmf_eagle} confirms that our method is able to recover
the PNSMFs properly.

\subsection{The growth of galaxies}%
\label{sub:the_growth_of_galaxies}

\begin{figure*}
    \centering
    \includegraphics[width=\linewidth]{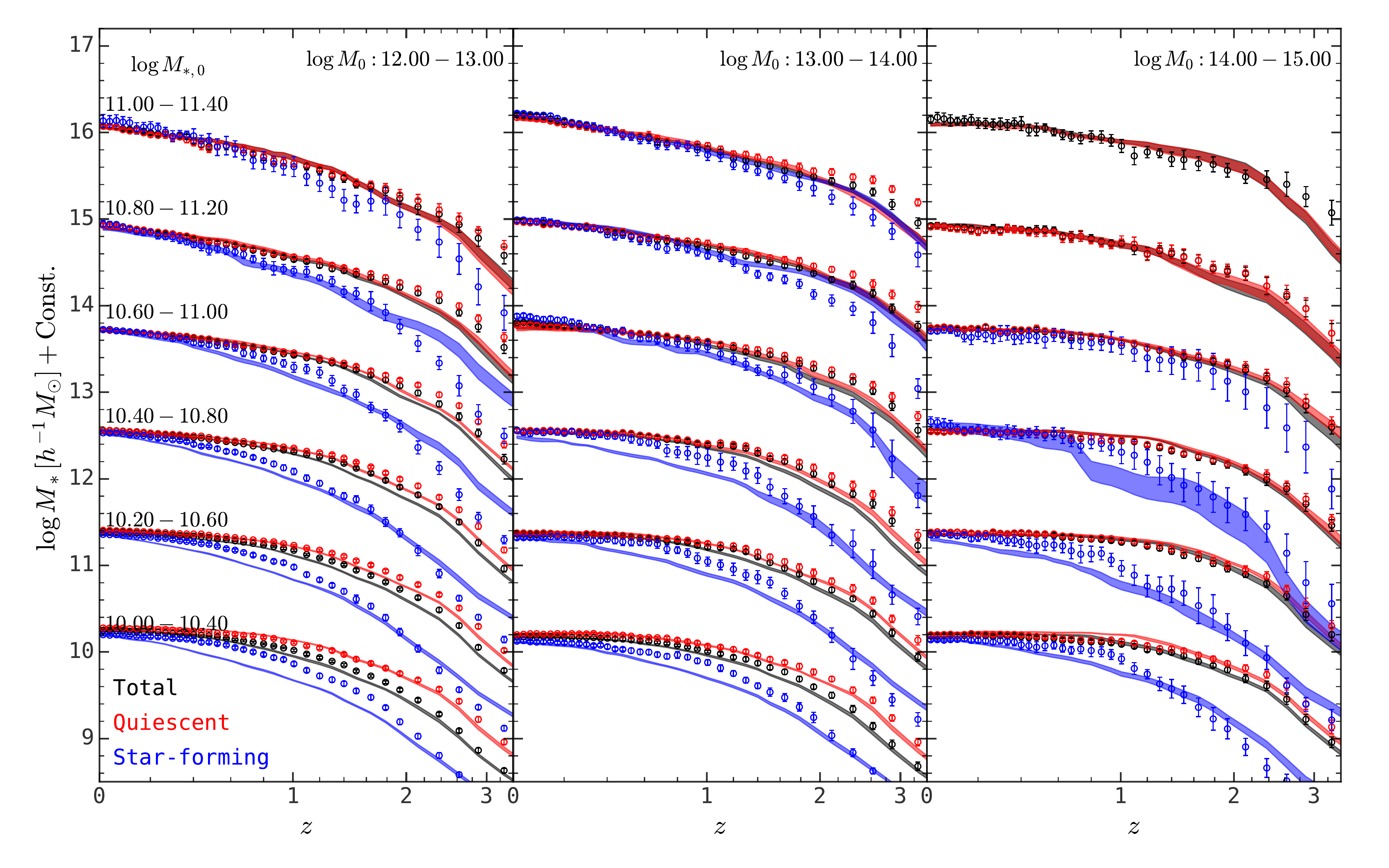}
    \caption{
        Median stellar mass evolution of main progenitor galaxies in the TNG
        simulation. Results are presented in bins of descendant stellar mass
        and descendant halo mass. Red/blue/black colors are for
        quiescent/star-forming/total populations of descendant galaxies. Filled
        regions are the results derived from galaxy merger trees in the TNG
        simulation. Symbols are the results of our method applied to the TNG
        simulation. All the errors are calculated using the bootstrap method.
    }%
    \label{fig:figure/main_progenitor_history_tng100}
\end{figure*}

Although tracing all the progenitor galaxies can provide a more complete
description of galaxy assembly histories, a simpler and yet illuminating
description can be obtained by examining how galaxies grow along their main
branches. Fig.\ \ref{fig:figure/main_progenitor_history_tng100} shows the
stellar mass evolution of the main progenitors for galaxies of different
present-day mass hosted by halos in three mass ranges, as indicated by $\log
M_0$ in each panel. The results for different descendant stellar mass bins are
shifted vertically for clarity. Shaded regions are obtained from galaxy merger
trees of the TNG simulation, and symbols are the predictions of our method.
Both errors are estimated using the bootstrap method. Results are shown
separately for the total, quiescent and star-forming populations of the
descendant galaxies, as black, red, and blue lines, respectively. It is clear
from the figure that our method can accurately reproduce the stellar mass
growth of the main progenitors for descendant galaxies of different stellar
masses and different host halo mass. Our method also works well for quiescent
descendants over the entire stellar mass and redshift ranges, and for
star-forming descendants of low-mass descendants at $z<2$. Our method
underestimates the progenitor stellar mass for massive star-forming galaxies at
$z\gtrsim 2$, especially in low-mass halos. This suggests that star formation
activities in massive galaxies are not fully determined by the assembly
histories of dark matter halos \citep[e.g.][]{chenHowEmpiricallyModel2021}. We
also test the performance of our method using the EAGLE simulation in
Appendix~\ref{sec:test_performance_on_eagle_simulation}, and
Fig.~\ref{fig:figure/main_progenitor_history_eagle} shows again that our method
works well in recovering the stellar mass growth of main progenitors in the
EAGLE simulation. Note that each of the descendant stellar mass bins used for
our presentations overlaps with its adjacent bins. This choice is to obtain
better statistics for each bin and provides more options to use the data. If
one is to use the data to constrain models, only data for independent mass bins
should be used.

\section{Applications to real observations}%
\label{sec:application}

The previous section shows that our method is capable of modeling the formation
histories of galaxies with different masses and star formation states. Here we
apply our method to observations to infer the stellar mass assembly histories
of galaxies in the real Universe.

\subsection{Observational data}%
\label{sub:observational_data}

\begin{figure}
    \centering
    \includegraphics[width=1\linewidth]{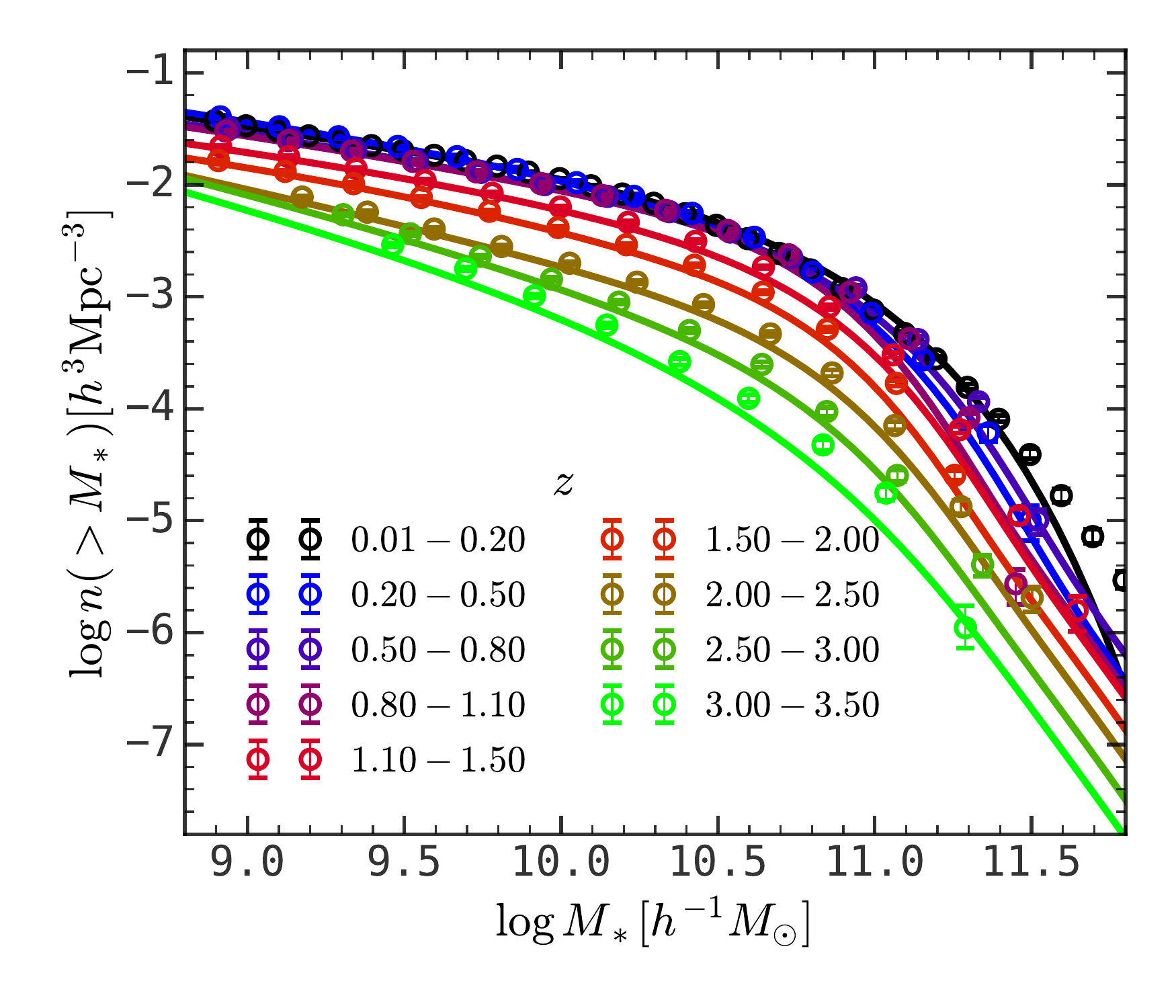}
    \caption{
        The cumulative stellar mass function for galaxies at $0.01 < z < 3.50$
        from the SDSS-GALEX and COSMOS2015 catalogues
        \citep{moustakasPRIMUSCONSTRAINTSSTAR2013,
        davidzonCOSMOS2015GalaxyStellar2017}. Circles with error bars are
        measurements, and solid lines are fitting results
        \citep{davidzonCOSMOS2015GalaxyStellar2017}.
    }%
    \label{fig:figure/smf}
\end{figure}

As shown in Section~\ref{sec:method}, assigning stellar masses to subhalos
requires stellar mass functions from $z=0$ to the redshift we want to probe.
Here we use the stellar mass function measurements from the SDSS-GALEX and
COSMOS2015 catalogues \citep{moustakasPRIMUSCONSTRAINTSSTAR2013,
davidzonCOSMOS2015GalaxyStellar2017}.\footnote{We note that the stellar mass
    function measurement in \citet{moustakasPRIMUSCONSTRAINTSSTAR2013} suffers
    from systematic effects in photometry and uses a stellar mass estimate that
    is different from \citet{davidzonCOSMOS2015GalaxyStellar2017}. To maintain
    consistency in observational data, we adopt the re-calibrated results in
    \citet{behrooziUniverseMachineCorrelationGalaxy2019} where an attempt was
made to deal with these issues.} Stellar masses are estimated using the
stellar population synthesis model of
\citet{bruzualStellarPopulationSynthesis2003}, the dust model of
\citet{calzettiDustContentOpacity2000}, and the initial mass function of
\citet{chabrierGalacticStellarSubstellar2003}. The cumulative stellar mass
functions obtained this way for galaxies at different redshift are
presented in Fig.~\ref{fig:figure/smf}.

\begin{figure}
    \centering
    \includegraphics[width=\linewidth]{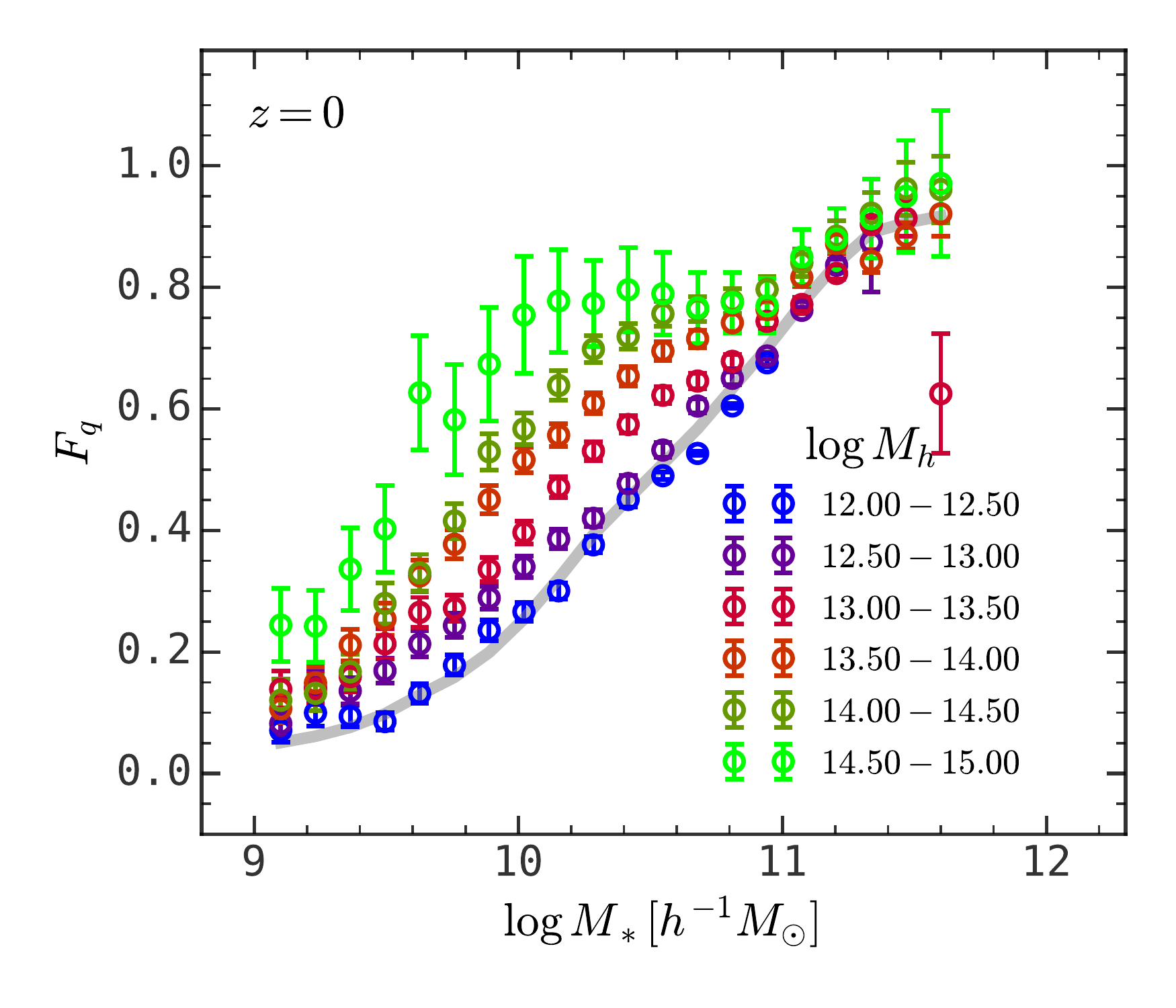}
    \caption{
        Quiescent fraction of galaxies at $0.01 < z < 0.20$ calculated from the
        MPA-JHU DR7 catalog \citep{brinchmannPhysicalPropertiesStarforming2004}
        and the group catalogue in \citet{yangGalaxyGroupsSDSS2007}. Circles
        with error bars are for galaxies with different host halo masses, and
        gray shaded regions are for all of the galaxies.
    }%
    \label{fig:figure/quenched_fraction}
\end{figure}

In order to assign star formation states to subhalos, we also need to know the
fraction of quiescent galaxies as a function of stellar mass and host halo
mass. The information of host halo mass can be obtained by grouping galaxies
into clusters and groups that can be used to represent host halos of galaxies.
At the present,  this can be done reliably only for the local Universe
\citep[e.g.][]{yangGalaxyGroupsSDSS2007, wangIdentifyingGalaxyGroups2020}.
Nevertheless, the data available at $z\sim 0$ are sufficient to study the
stellar mass evolution conditioned on descendant properties at $z\sim 0$, as is
done in \S~\ref{sec:performance}. Here we use the criterion in
equation~(\ref{eq:q}) to separate quiescent galaxies from their star-forming
counterparts, where the star formation rate measurements come from MPA-JHU DR7
\footnote{\url{http://www.mpa-garching.mpg.de/SDSS/DR7/}}
\citep{brinchmannPhysicalPropertiesStarforming2004}, and the quiescent fraction
as a function of stellar mass and host halo mass is presented in
Fig.~\ref{fig:figure/quenched_fraction}. We note that the quiescent fractions
have been previously obtained with the SDSS galaxy sample
\citep[e.g.][]{Wetzel2012, Hirschmann2014, wangELUCIDIVGalaxy2018,
Li2020quenchedfractions}, but different studies differ to varying degrees,
particularly at low stellar mass and high halo mass due to different criteria
to separate quiescent and star-forming populations. It is thus necessary to
apply the same criterion for both observational measurements and theoretical
models for a meaningful comparison, as the case in the present work. Tests show
that we're able to reproduce the measurements in the literature if exactly the
same criteria are adopted.

To apply our method to observational data, we first assign stellar mass to
subhalos at nine snapshots following the procedures in
\S\,\ref{sub:assigning_stellar_mass_to_subhalos}, where each snapshot is
selected to match the middle value of the redshift interval in which the
stellar mass function is measured from observational data (see
Fig.~\ref{fig:figure/smf}). We then apply the method in
\S\,\ref{sub:assigning_star_forming_state} to assign star formation states to
subhalos at $z\sim 0.1$ using the observed quiescent fraction shown in
Fig.~\ref{fig:figure/quenched_fraction}. Finally, we calculate the distribution
and redshift evolution of the progenitor stellar mass for descendant galaxies
of different stellar mass, halo mass, and star formation states. The results
are presented in subsequent subsections. Note that the lowest redshift we can
probe is $z\sim 0.1$, since $M_{*, 0}$ and $M_0$ used in observational results
are actually measured at $z\sim 0.1$.

\subsection{Stellar mass evolution since $z\sim 3$}%
\label{sub:stellar_mass_evolution_history_to_zsim_3_}

\begin{figure*}
    \centering
    \includegraphics[width=0.9\linewidth]{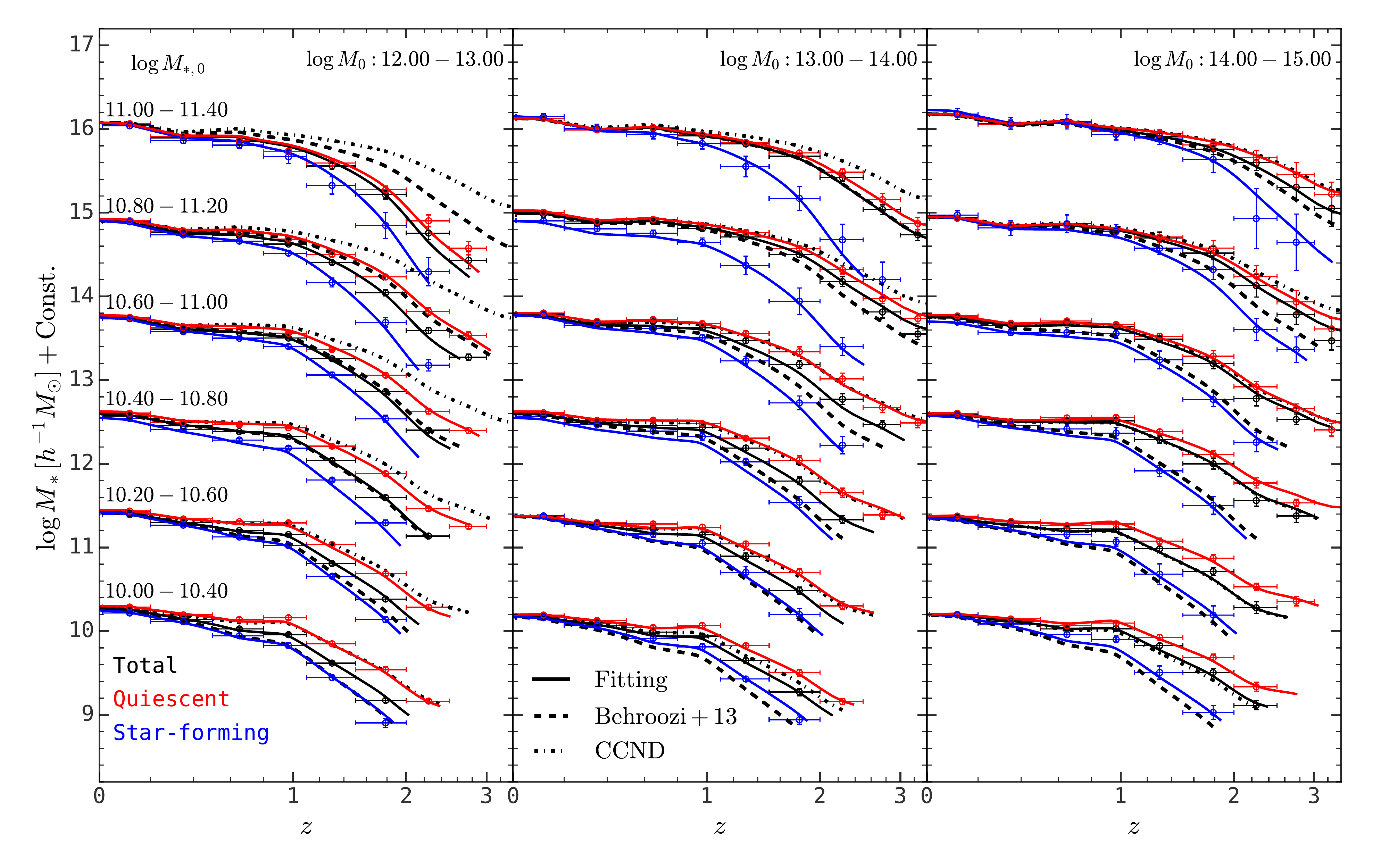}
    \caption{
        Median stellar mass evolution of main progenitor galaxies derived from
        the observational data with our method. Results are presented in bins
        of descendant stellar mass, $M_{*, 0}$, and descendant halo mass,
        $M_0$. Red/blue/black colors are for quiescent/star-forming/total
        populations of descendant galaxies. Circles are the direct results of
        our method, with horizontal error bars showing the corresponding
        redshift bins and vertical error bars showing the standard deviations
        estimated with the bootstrap method. Solid lines are results derived
        from our linear fitting to the evolution of the cumulative number
        density (See equation~(\ref{eq:slope})). Dashed lines are the results
        in \citet{behrooziUSINGCUMULATIVENUMBER2013}. Dash-dotted lines are the
        results derived with the constant cumulative number density (CCND)
        method \citep[see][]{vandokkumGROWTHMASSIVEGALAXIES2010}.
    }%
    \label{fig:figure/main_progenitor_history_observation}
\end{figure*}

\begin{figure}
    \centering
    \includegraphics[width=0.9\linewidth]{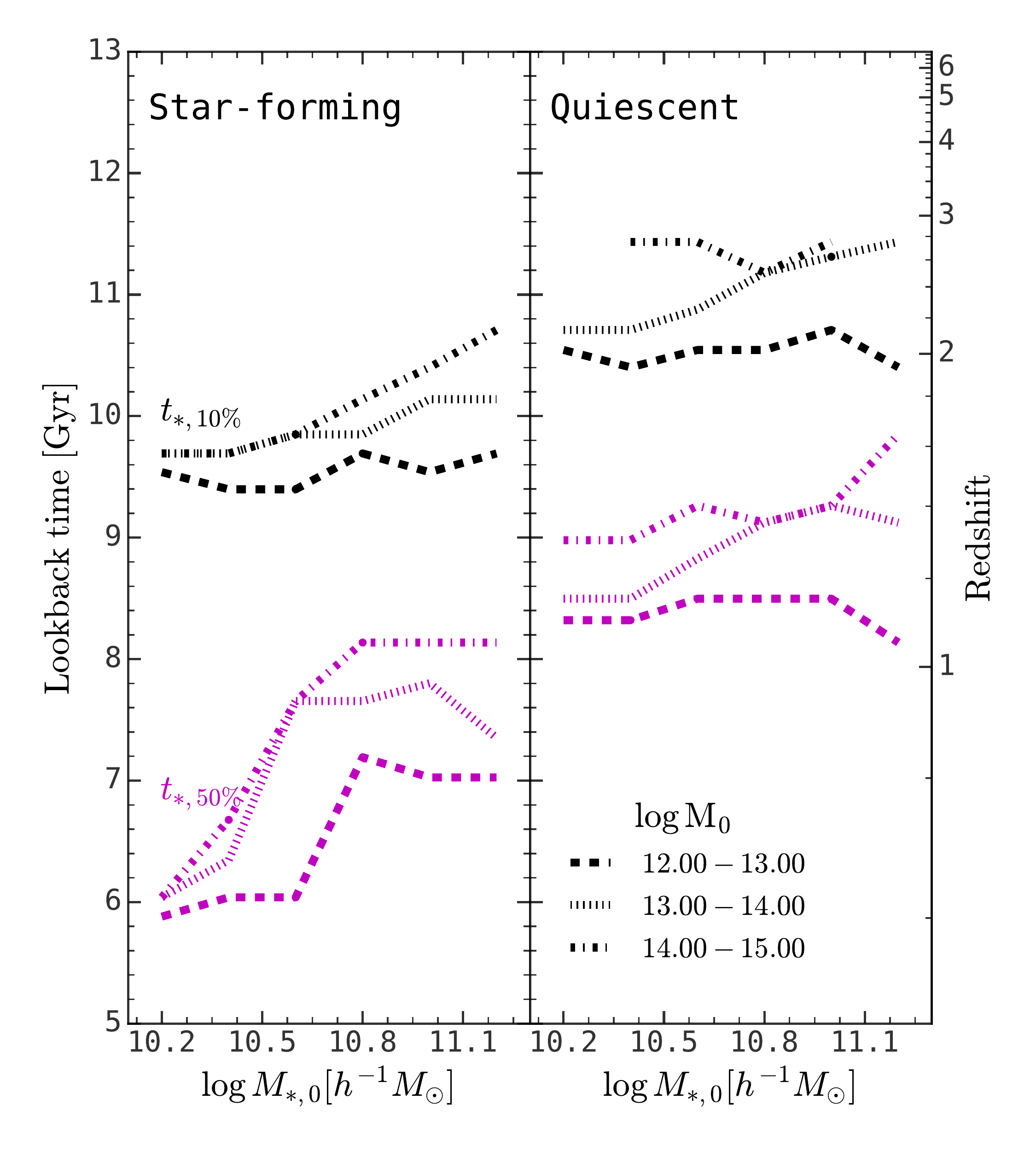}
    \caption{
        Galaxy formation time defined as the lookback time that 10\% ({\it black})
        and 50\% ({\it magenta}) of the final stellar mass is assembled into the main
        progenitor. Results are presented for star-forming ({\it left panel}) and
        quiescent ({\it right panel}) descendant galaxies with different
        stellar mass and halo mass.
    }%
    \label{fig:figure/galaxy_formation_time_observation}
\end{figure}

\begin{figure*}
    \centering
    \includegraphics[width=0.9\linewidth]{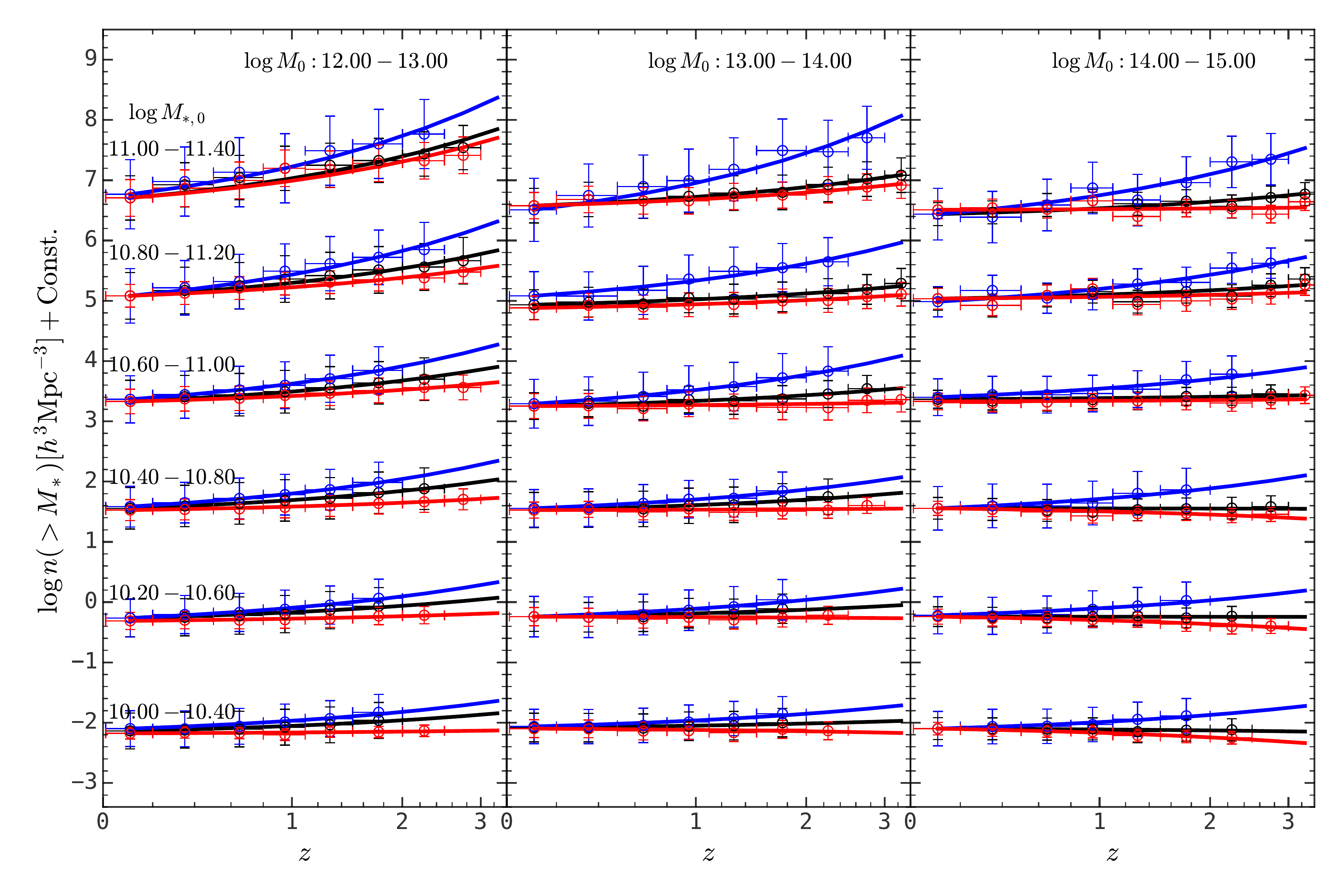}
    \caption{
        Median cumulative number density evolution of main progenitor galaxies
        derived from the observational data with our method. Results are
        presented in bins of descendant stellar mass, $M_{*, 0}$, and
        descendant halo mass, $M_0$. Red/blue/black colors are for
        quiescent/star-forming/total populations of descendant galaxies.
        Circles with error bars are derived from our method, where the
        horizontal error bars show the redshift bins, and the vertical error
        bars are derived with the bootstrap method. Solid lines are linear
        fitting results (See equation~(\ref{eq:slope})). Note that this figure
        can be converted into the solid lines and symbols in Fig.\
        \ref{fig:figure/main_progenitor_history_observation} using the
        cumulative stellar mass function shown in Fig.\ \ref{fig:figure/smf}.
    }%
    \label{fig:figure/main_progenitor_history_cnd_observation}
\end{figure*}

\begin{figure*}
    \centering
    \includegraphics[width=\linewidth]{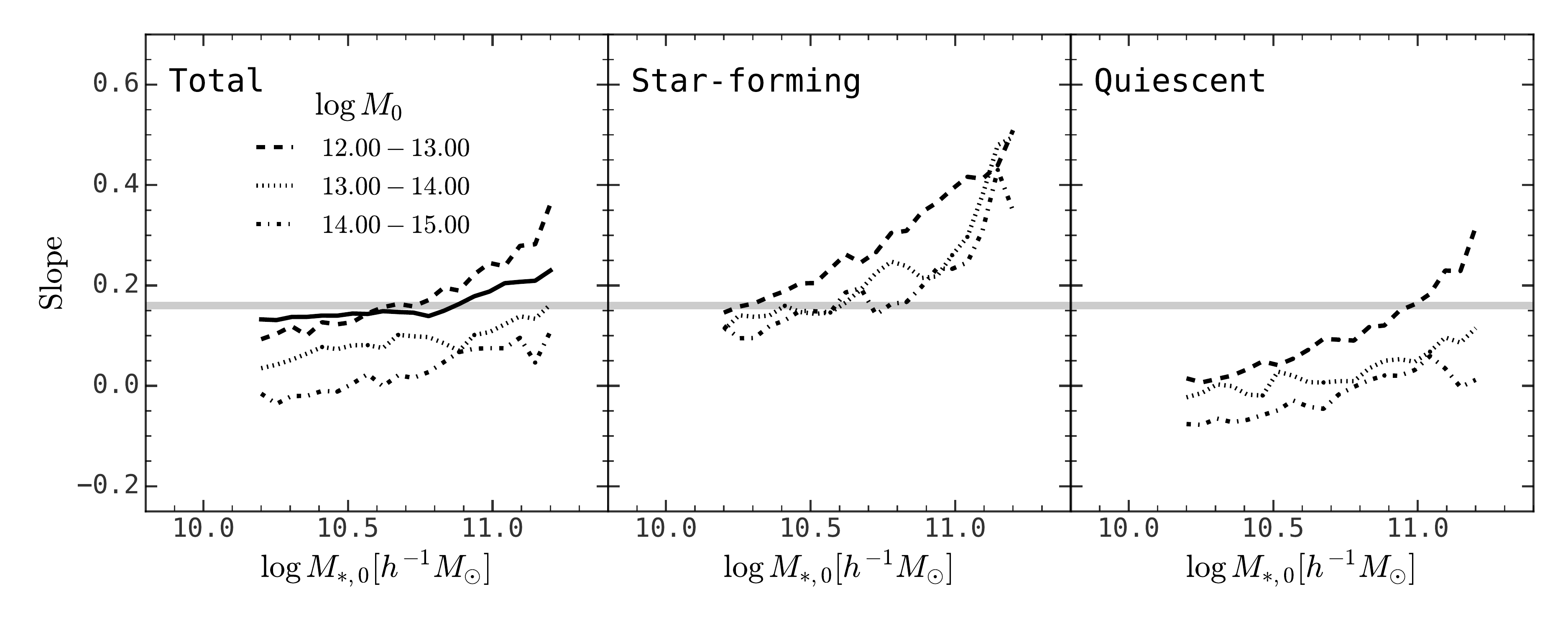}
    \caption{
        The slope of the redshift evolution in the cumulative number density
        (See equation~(\ref{eq:slope})) and its dependence on descendant
        stellar mass, descendant halo mass, and descendant star formation
        state. The horizontal line is the result obtained in
        \citet{behrooziUSINGCUMULATIVENUMBER2013} using the evolving cumulative
        number density method, i.e. \texttt{Slope} = 0.16, which neglects the
        dependence on descendant halo mass and descendant star formation state.
        As a check, we also implement the same evolving cumulative number
        density method in \citet{behrooziUSINGCUMULATIVENUMBER2013} and the
        result is shown as the black solid line in the first panel, which is
        consistent with the horizontal line.
    }%
    \label{fig:figure/cnd_slope_dependence}
\end{figure*}

Circles in Fig.~\ref{fig:figure/pnsmf_observation_zstarve1} show PNSMFs
obtained from our observational data, with error bars obtained using the bootstrap
method. The overall trends of the observational results are similar to those
given by the TNG simulation, except for some noticeable discrepancies at high
$z$ for descendants of low masses. These discrepancies are expected, as the TNG
simulation over-predicts the stellar mass function at $1.0\lesssim z \lesssim
3.0$ \citep{pillepichFirstResultsIllustrisTNG2018}.

Fig.~\ref{fig:figure/main_progenitor_history_observation} shows the median
stellar mass evolution of main progenitor galaxies for descendants with
different stellar mass, halo mass, and star formation state. Circles show
the results obtained by applying our method to the observational data, where
the horizontal error bars show the redshift bins and the vertical error bars
show the standard deviation calculated from bootstrap samples.

We can see that the progenitors of descendant galaxies of a given stellar mass
are less massive in lower-mass halos, as shown by the black symbols in
Fig.~\ref{fig:figure/main_progenitor_history_observation}, indicating that
these galaxies evolve more rapidly over the redshift range covered. For
example, consider the stellar mass growth from $z\sim 2$ to $z=0$. For massive
galaxies with $M_{*, 0}\sim 10^{11}h^{-1}\rm M_{\odot}$, those residing in
halos with $M_0 \sim 10^{12}$-$10^{13}h^{-1}\rm M_{\odot}$ have increased their
stellar mass by a factor of $\sim 7$, while those with $M_0 \gtrsim 10^{14}\rm
M_{\odot}$ have increased only a factor of $\sim 2.5$. For lower-mass galaxies
with $M_{*, 0}\sim 10^{10}h^{-1}\rm M_{\odot}$, the stellar mass growth factors
for galaxies residing in low-mass and massive halos are $\sim 12$ and $\sim 5$,
respectively. Roughly speaking, galaxies that end up in low-mass descendant
halos grow $\sim 2.5$ times as fast as those that end up in massive halos.

In general, for similar descendant stellar mass and descendant halo mass,
star-forming galaxies at $z\sim 0$ have progenitors that are less massive than
those of their quiescent counterparts, indicating that they grow more rapidly.
For example, between $z\sim 1.8$ and $z\sim 0$, low-mass star-forming galaxies
grow faster than their quiescent counterparts by a factor of $\sim 4$, while
this factor is $\sim 2$ for massive descendant galaxies. Both factors depend
only weakly on the host halo mass. These results indicate that the dependence of
stellar mass growth on descendant halo mass is mainly due to the fact that the
fraction of quiescent galaxies is higher in more massive halos at $z\sim 0$.

We can also quantify the stellar mass growth by identifying the epoch when a
galaxy has obtained a certain fraction of its final stellar mass.
Fig.~\ref{fig:figure/galaxy_formation_time_observation} shows the lookback time
when the main progenitor has accumulated 10\% ($t_{*, 10\%}$) and 50\% ($t_{*,
50\%}$) of its final stellar mass, respectively. Results are shown for
star-forming and quiescent descendant galaxies with different stellar masses and
host halo masses. Overall, quiescent descendant galaxies formed earlier than
their star-forming counterparts by $\sim 1-2$ Gyr depending on the descendant
stellar mass and descendant halo mass. It is clear that galaxies with the same
stellar mass but hosted by more massive descendant halos formed earlier than
the ones that end up in low-mass halos, and this halo mass dependence appears
to be stronger at higher stellar mass. The only exception is that the halo mass
dependence is rather weak at the lowest stellar mass bin, where star-forming galaxies
in massive halos appear to be as young as their counterparts in low-mass halos.
Some of the galaxies in this population may be accreted into massive halos only
recently, and their evolution may not yet have been affected by the cluster
environment \citep[e.g.][]{wetzelGalaxyEvolutionGroups2013}.

The general trends presented above are consistent with the results in both the
TNG and EAGLE simulations (See
Figs.~\ref{fig:figure/main_progenitor_history_tng100} and
\ref{fig:figure/main_progenitor_history_eagle}). They are not revealed in
previous methods because these methods neglect the dependence of the stellar
mass evolution on descendant halo mass and descendant star formation state.

\begin{figure*}
    \centering
    \includegraphics[width=0.9\linewidth]{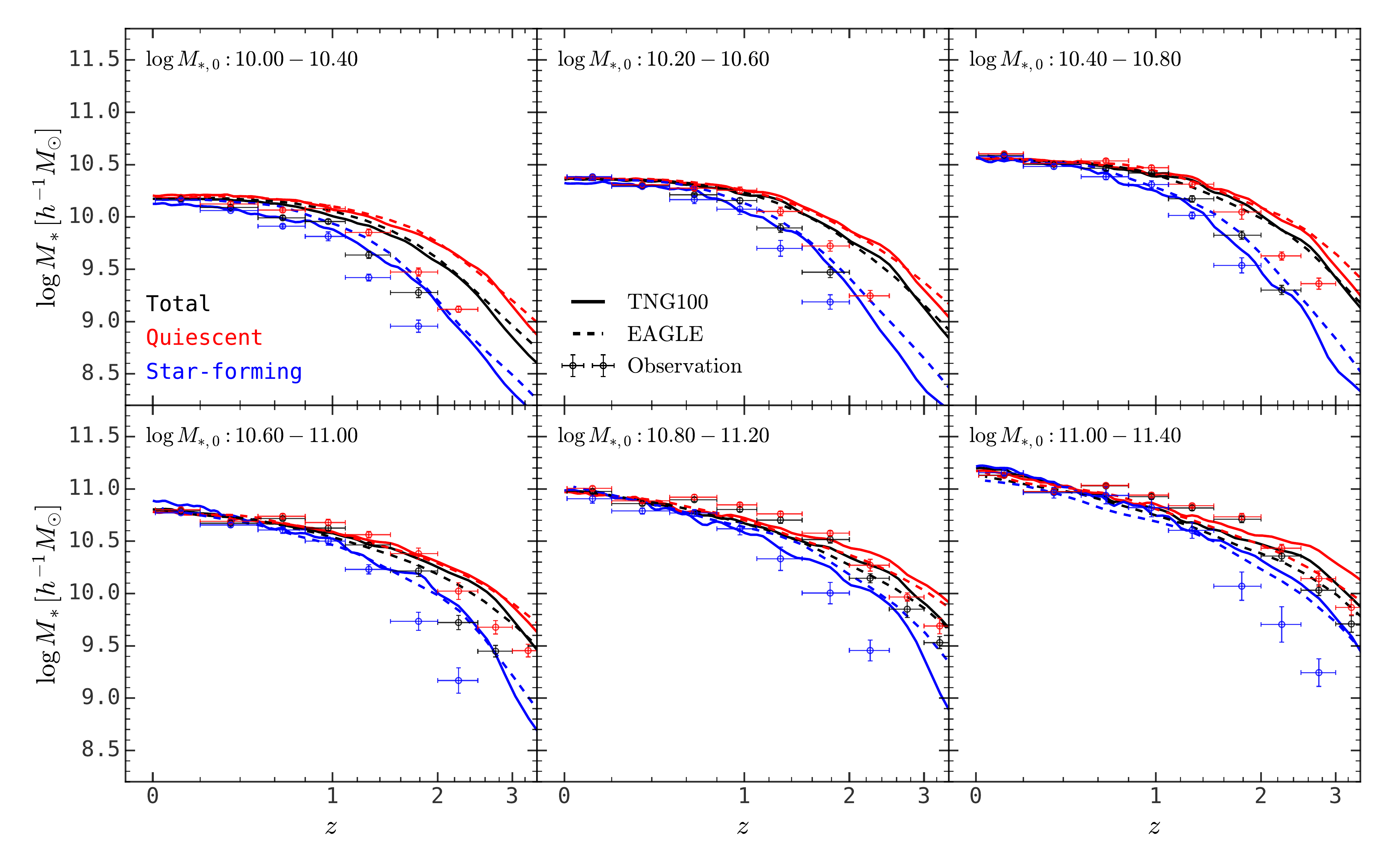}
    \caption{
        Comparison of the median stellar mass evolution of main progenitor
        galaxies obtained by applying our method to simulations and
        observation. Here we only present results for galaxies with descendant
        halo mass of $13 \leq \log(M_0/[h^{-1}M_{\odot}]) < 14$ and omit results
        in other mass ranges since they are very similar. Circles with error
        bars are derived from observational data. Solid and dashed lines are
        obtained from galaxy merger trees in the TNG and EAGLE simulations,
        respectively. Red/blue/black colors are for
        quiescent/star-forming/total populations of descendant galaxies.
    }%
    \label{fig:figure/main_progenitor_history_compare_sim_obs}
\end{figure*}

In the simple formalism described in \citet{behrooziUSINGCUMULATIVENUMBER2013},
the stellar mass evolution of main progenitors is followed by using the
cumulative number density as an equivalent of the stellar mass:
\begin{equation}
    \log \left(\frac{n_z(>M_*)}{n_0(>M_{*, 0})}\right) = {\tt Slope}\times z,
    \label{eq:slope}
\end{equation}
where the cumulative number density is defined as
\begin{equation}
    n_z(>M_*) = \int_{M_*}^{\infty}\Phi_z(M_*^{\prime})dM_*^{\prime},
\end{equation}
with $\Phi_z(M_*)$ being the stellar mass function at redshift $z$. The
observed relations between $M_*$ and $n(> M_*)$ at different redshift are shown
in Fig.~\ref{fig:figure/smf}. In this way, one can use a single parameter, the
\texttt{Slope} defined in equation~(\ref{eq:slope}), to describe the main
progenitor stellar mass evolution for given descendant galaxies. Using the
cumulative number density to replace the stellar mass has the advantage that it
is less affected by systematic uncertainties in the stellar mass estimate
\citep{conroyModelingPanchromaticSpectral2013}.

Fig.~\ref{fig:figure/main_progenitor_history_cnd_observation} shows the
redshift evolution of the cumulative number density of main progenitor
galaxies. Each evolution trajectory is fitted with a linear function given by
equation~(\ref{eq:slope}), and shown as solid lines. We emphasize again that
the evolution of the cumulative number density is mathematically equivalent to
the evolution in stellar mass. The advantage of going through an intermediate
step of using the cumulative number density is that it can be described well by
the simple linear function, equation~(\ref{eq:slope}), specified by a {\tt
slope}. Overall, the cumulative number density increases faster with increasing
$z$ for star-forming galaxies than their quiescent counterparts, implying that
quiescent galaxies have more massive progenitors than star-forming galaxies.

To see this more clearly, we plot in Fig.~\ref{fig:figure/cnd_slope_dependence}
the slope defined in equation~(\ref{eq:slope}) as a function of descendant
properties. For reference, the black solid line in the left panel shows the
result obtained by using the evolving number density method of
\citet{behrooziUSINGCUMULATIVENUMBER2013}. In this case, the dependence on the
descendant stellar mass, $M_{*, 0}$, is weak, consistent with the result of
\citet{behrooziUSINGCUMULATIVENUMBER2013}, shown with the gray horizontal line.
Inspecting the dependence on both the descendant stellar mass, $M_{*, 0}$, and
the descendant halo mass, $M_0$, we see that the slope parameter has a modest
dependence on host halo mass: at fixed stellar mass the galaxies in massive
halos have a smaller slope, and thus have evolved more slowly. The middle and
right panels show results separately for the star-forming and quiescent
descendants. It is clear that, on average, the slope parameter for quiescent
descendant galaxies is smaller than their star-forming counterparts, indicating
a slower growth of stellar mass in quiescent galaxies over the redshift range
being considered. Moreover, the dependence on the host halo mass in the middle
and right panels is somewhat weaker than that in the left panel, indicating
that the dependence on the host halo mass in the left panel is partly caused by
the dependence of the quiescent fraction on the host halo mass. These results
also show that, in order to recover the details of the stellar mass growth of
galaxies, it is important to control the star formation and halo properties of
galaxies in addition to the stellar mass of galaxies
\citep[e.g.][]{clauwensLargeDifferenceProgenitor2016}.

To show the difference between our results and previous ones more clearly, we
plot our fitting results as solid lines in
Fig.~\ref{fig:figure/main_progenitor_history_observation} using the relation
between $M_*$ and $n(>M_*)$ shown in Fig.~\ref{fig:figure/smf}. In addition, we
also plot the results of \citet{behrooziUSINGCUMULATIVENUMBER2013}
(\texttt{Slope} = 0.16) as dashed lines, and results of the constant cumulative
number density (CCND) method (\texttt{Slope} = 0) as dashed-dotted lines in
Fig.~\ref{fig:figure/main_progenitor_history_observation}. Our method predicts
more massive main progenitor galaxies for quiescent descendants than for their
star-forming counterparts. In addition, our method also predicts more massive
main progenitors for descendants in massive halos. This is expected because
stellar mass growth is suppressed in dense regions at the present day. Both of
these trends are also found in the TNG and EAGLE simulations as shown below,
but they are not recovered by previous methods since they do not consider the
dependence of the stellar mass growth on the host halo mass and star formation
state of descendant galaxies.

\subsection{Comparison between simulations and observation}%
\label{sub:comparison_between_simulations_and_observation}

In Fig.~\ref{fig:figure/main_progenitor_history_compare_sim_obs}, we compare
the stellar mass evolution of main progenitor galaxies obtained from
observational data with results in the TNG and EAGLE simulations. Here we only
show the comparison in the descendant halo mass bin of $13 < \log (M_0 /
[h^{-1} M_{\odot}]) < 14$ since other bins have similar trends. For massive
descendant galaxies with $M_{*, 0}>10^{11}M_\odot$, both simulations can reproduce the
evolution trend for the \texttt{quiescent} and \texttt{total} populations of
descendant galaxies. However, there is a discrepancy for
\texttt{star-forming} descendants at $z\gtrsim 1.5$. This discrepancy arises
because both simulations over-predict the stellar mass function at high-$z$
\citep{pillepichFirstResultsIllustrisTNG2018,
furlongEvolutionGalaxyStellar2015}. The two simulations also
over-predict the progenitor stellar mass for low-mass descendants regardless
of their star formation states, again because they overestimate the
stellar mass function at high $z$.

\section{Summary}
\label{sec:summary}

Connecting progenitor and descendant galaxies across cosmic time statistically
has proven to be a promising way to investigate galaxy evolution
\citep[e.g.][]{brownEvolvingLuminosityFunction2007,
    vandokkumGROWTHMASSIVEGALAXIES2010, vandevoortGalaxyGrowthRedshift2016,
    behrooziUSINGCUMULATIVENUMBER2013, lejaTracingGalaxiesCosmic2013,
jaacksCONNECTINGDOTSTRACKING2016, hillMassColorStructural2017}. Previous
investigations only used the stellar mass of galaxies to establish this
connection, neglecting the dependence on other properties, such as star
formation activities and host halo mass of descendant galaxies. In the present
paper, we propose a general framework, which leverages galaxy-halo connections
that have been established in the past, to connect galaxies across cosmic time
using subhalo merger trees expected from the current model of structure
formation. Our main results are summarized as follows:
\begin{enumerate}

    \item We tested our method by applying it to the TNG and EAGLE cosmological
        simulations. Our results show that our method can properly recover the
        progenitor stellar mass distribution up to $z\sim 3$ for star-forming
        and quiescent descendant galaxies with different stellar masses and host
        halo masses. Our method can also recover the stellar mass growth along
        the main branches of galaxies.

    \item Applying our method to observational data, we derived the
        stellar mass assembly histories since $z\sim 3$ for galaxies in the
        local Universe, including the stellar mass evolution along the main
        branch.

    \item We found that the stellar mass growth depends strongly on the
        descendant halo mass at $z=0$. Since $z\sim 1.8$, massive descendant
        galaxies in low-mass halos have increased their stellar mass by a
        factor of $\sim 7$, while low-mass descendant galaxies have increased
        their masses by a factor of $\sim 12$. In massive descendant halos, the
        stellar mass growth for massive and low-mass descendant galaxies is
        about factors of 2.5 and 5, respectively. Thus, galaxies that end up in
        present low-mass groups accumulate their stellar mass $\sim 2.5$ times
        as fast as those that end up in massive clusters.

    \item We found that the stellar mass growth also depends on the current star
        formation activity of descendant galaxies. From $z\sim 2$ to $z\sim
        0$, the stellar mass growth rate for star-forming descendant galaxies
        is about $2\sim 4$ times faster than their quiescent counterparts,
        quite independent of descendant halo mass. This suggests that the
        dependence on the host halo mass is mainly driven by the correlation
        between the quiescent fraction and the host halo mass.

    \item We found that the stellar mass evolution can be well described by a
        linear function in the logarithm of the cumulative number density, as
        in equation (\ref{eq:slope}), for descendant galaxies with given
        stellar mass, star-formation state, and host halo mass. The resulting
        parameter, $\texttt{Slope}$, also shows a strong dependence on
        both the descendant star formation state and descendant halo mass.
        The dependence on the descendant halo mass becomes weaker when the
        descendant star formation state is fixed. These trends are consistent
        with the analysis using stellar mass alone.

    \item We compared the stellar mass evolution derived from observation with
        results in the TNG and EAGLE simulations. We found that both
        simulations over-predict the progenitor stellar mass for low-mass
        descendant galaxies, because both simulations over-predict the stellar
        mass functions at high $z$.

\end{enumerate}

Again, it is important to take into account the potential dependencies of the
evolution on descendant star formation activities and descendant halo mass, and
the method proposed here provides a more accurate framework for such modeling.

\section*{Acknowledgements}

The authors thank the anonymous referee for their helpful comments that
improved the quality of the manuscript. This work is supported by the National
Key R\&D Program of China (grant No. 2018YFA0404502, 2018YFA0404503), and the
National Science Foundation of China (grant Nos. 11821303, 11973030, 11673015,
11733004, 11761131004, 11761141012). The authors acknowledge the Tsinghua
Astrophysics High-Performance Computing platform at Tsinghua University for
providing computational and data storage resources that have contributed to the
research results reported within this paper.

Funding for the SDSS and SDSS-II has been provided by the Alfred P. Sloan
Foundation, the Participating Institutions, the National Science Foundation,
the U.S. Department of Energy, the National Aeronautics and Space
Administration, the Japanese Monbukagakusho, the Max Planck Society, and the
Higher Education Funding Council for England. The SDSS Web Site is
http://www.sdss.org/.

The SDSS is managed by the Astrophysical Research Consortium for the
Participating Institutions. The Participating Institutions are the American
Museum of Natural History, Astrophysical Institute Potsdam, University of
Basel, University of Cambridge, Case Western Reserve University, University of
Chicago, Drexel University, Fermilab, the Institute for Advanced Study, the
Japan Participation Group, Johns Hopkins University, the Joint Institute for
Nuclear Astrophysics, the Kavli Institute for Particle Astrophysics and
Cosmology, the Korean Scientist Group, the Chinese Academy of Sciences
(LAMOST), Los Alamos National Laboratory, the Max-Planck-Institute for
Astronomy (MPIA), the Max-Planck-Institute for Astrophysics (MPA), New Mexico
State University, Ohio State University, University of Pittsburgh, University
of Portsmouth, Princeton University, the United States Naval Observatory, and
the University of Washington.

\section*{Data availability}

The data underlying this article will be shared on reasonable request to the
corresponding author. The computation in this work is supported by the HPC
toolkit \specialname[hipp] at \url{https://github.com/ChenYangyao/hipp}.

\bibliographystyle{mnras}
\bibliography{bibtex.bib}

\appendix

\section{Scatter of stellar mass-halo mass relation in TNG}%
\label{sec:scatter_of_stellar_mass_halo_mass_relation_in_tng300_1}

\begin{figure*}
    \centering
    \includegraphics[width=1\linewidth]{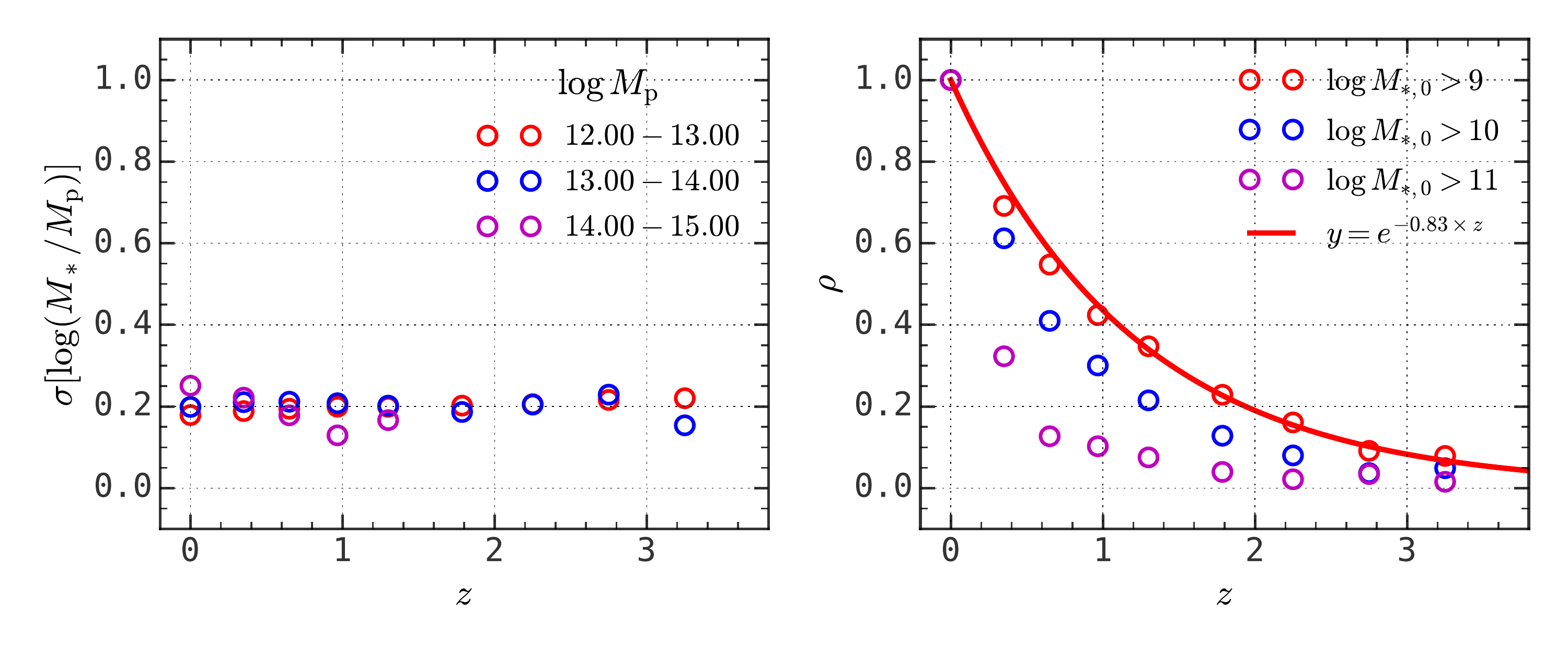}
    \caption{
        The left panel shows the scatter of the stellar mass-halo mass ratio,
        $r$, in TNG simulation from $z=3$ to $z=0$, in bins of peak halo mass
        at $z=0$, $M_{\rm p}$. The right panel shows the Pearson correlation
        coefficient defined in equation~(\ref{eq:pearson}), with the red solid
        curve showing the fitting result for $M_{*, 0}> 10^9M_{\odot}$.
}%
    \label{fig:figure/sm_hm_relation_in_TNG}
\end{figure*}

\begin{figure*}
    \centering
    \includegraphics[width=0.9\linewidth]{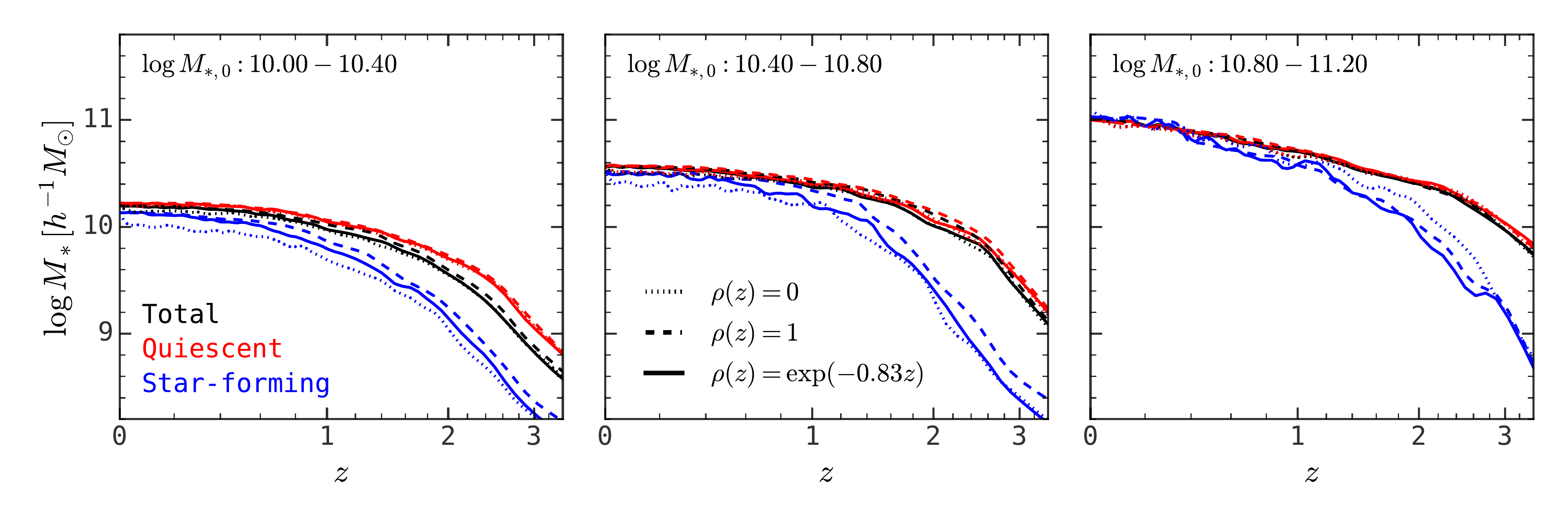}
    \caption{
        Median stellar mass evolution of main progenitor galaxies for
        descendant galaxies with $13\leq \log M_0/[h^{-1}M_{\odot}] < 14$,
        predicted by applying our method to the TNG simulation. Dotted lines
        show results assuming $\rho(z) = 0$, i.e. the scatter of the
        $M_*-M_{\rm p}$ relation is uncorrelated between adjacent snapshots.
        Dashed lines show results assuming $\rho(z) = 1$, i.e. the scatter of
        the $M_*-M_{\rm p}$ relation is fully determined by the descendant mass
        at $z=0$. The solid lines shows results with $\rho(z) = \exp(-0.83z)$,
        which is calibrated using the TNG simulation (see
        Fig.~\ref{fig:figure/sm_hm_relation_in_TNG}).
    }%
    \label{fig:figure/dependence_on_rho_z}
\end{figure*}

The left panel of Fig.\,\ref{fig:figure/sm_hm_relation_in_TNG} shows the
scatter of $\log(M_*/M_{\rm p})$ as function of redshift in bins of $M_{\rm p}$
from the TNG simulation. At each snapshot, we calculate the stellar mass to
peak halo mass ratio for all galaxies and calculate the standard deviation in
three peak halo mass bins, where the stellar mass and peak halo mass for TNG
galaxies are defined in \S\,\ref{sub:data_description}. It can be seen that the
scatter does not depend on redshift or $M_{\rm p}$. The right panel of
Fig.\,\ref{fig:figure/sm_hm_relation_in_TNG} shows the evolution of the Pearson
correlation coefficient, i.e.
\begin{equation}
    \rho(z) = \frac{\sum_i (r_{z, i} - \bar r_z)(r_{0, i}-\bar r_0)}{\sqrt{\sum_i(r_{z, i} - \bar r_z)^2\times \sum_j(r_{0, j} - \bar r_0)^2}},
    \label{eq:pearson}
\end{equation}
where $r_{0, i} = \log (M_{*, 0, i}/M_{{\rm d}, 0, i})$ is for galaxies at
$z=0$, $r_{z, i} = \log (M_{*, z, i} / M_{{\rm d}, z, i})$ is for the main
progenitor galaxies at the redshift of $z$. Here $M_{{\rm d}, z, i}$ is the
tentative stellar mass evaluated with the SHAM method (see
equation~(\ref{eq:tentative})), and $M_{{\rm d}, z, i}$ is the value at $z=0$.
We fit the evolution of the correlation coefficient for $M_{*, 0}
> 10^9M_{\odot}$ with an exponential function, as shown by the red solid
line and the annotation in the panel.

Fig.~\ref{fig:figure/dependence_on_rho_z} shows the stellar mass evolution
histories produced by applying our method to the TNG simulation with different
choices of $\rho(z)$. As one can see, our results are insensitive to the
specific choice of $\rho(z)$. From equations~(\ref{eq:tentativa_0}) and
(\ref{eq:tentative}) one can see that both $\epsilon_0$ and $\epsilon_z$ are
random variables with means equal to zero. Thus, the effect of $\rho(z)$ is
expected to be small as long as the sample of stellar mass histories is
sufficiently large.

\section{Test performance on EAGLE simulation}%
\label{sec:test_performance_on_eagle_simulation}

\begin{figure*}
    \centering
    \includegraphics[width=1\linewidth]{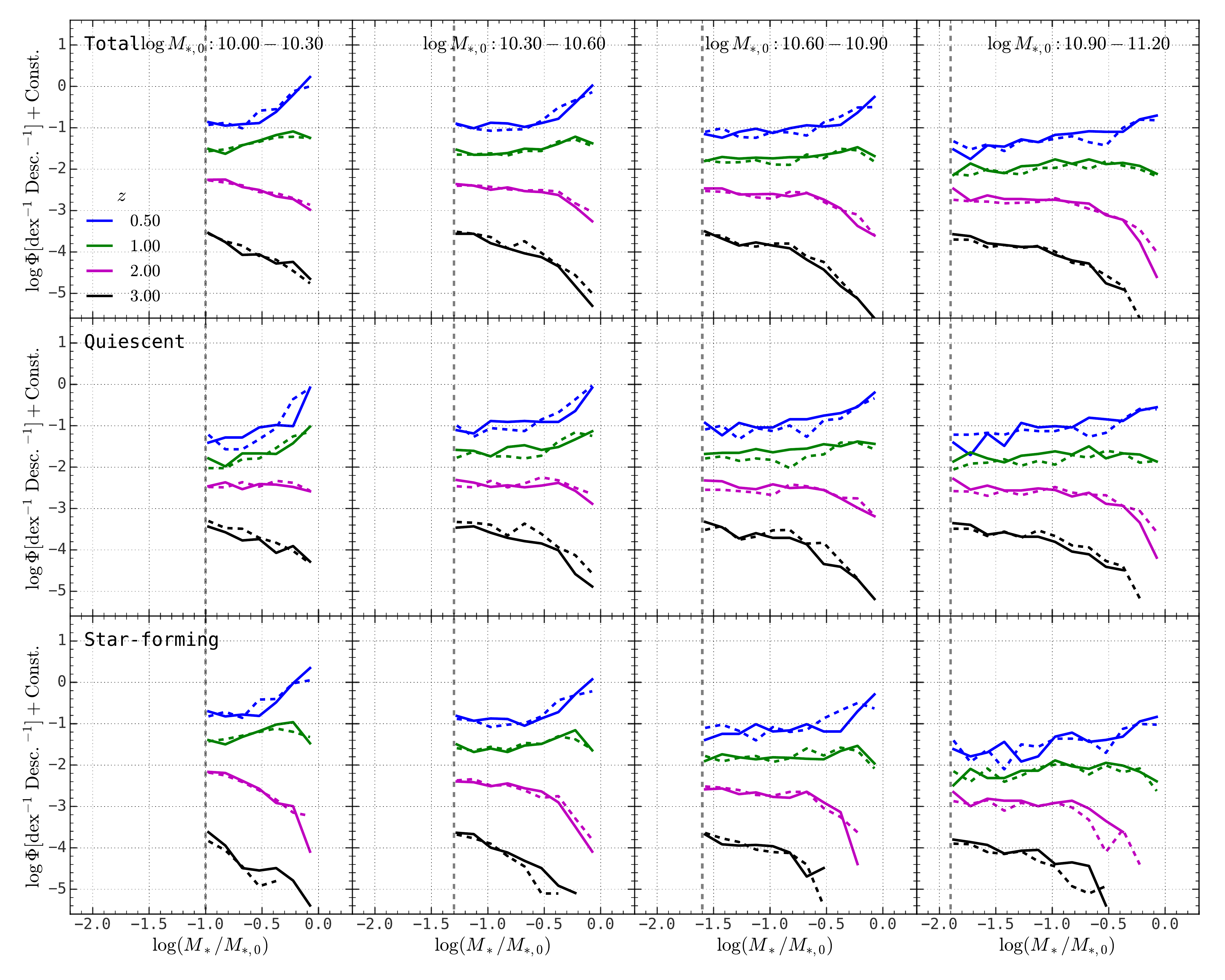}
    \caption{
        Similar to Fig.~\ref{fig:figure/pnsmf_observation_zstarve1}, but for
        the EAGLE simulation.
    }%
    \label{fig:figure/pnsmf_eagle}
\end{figure*}

\begin{figure*}
    \centering
    \includegraphics[width=1\linewidth]{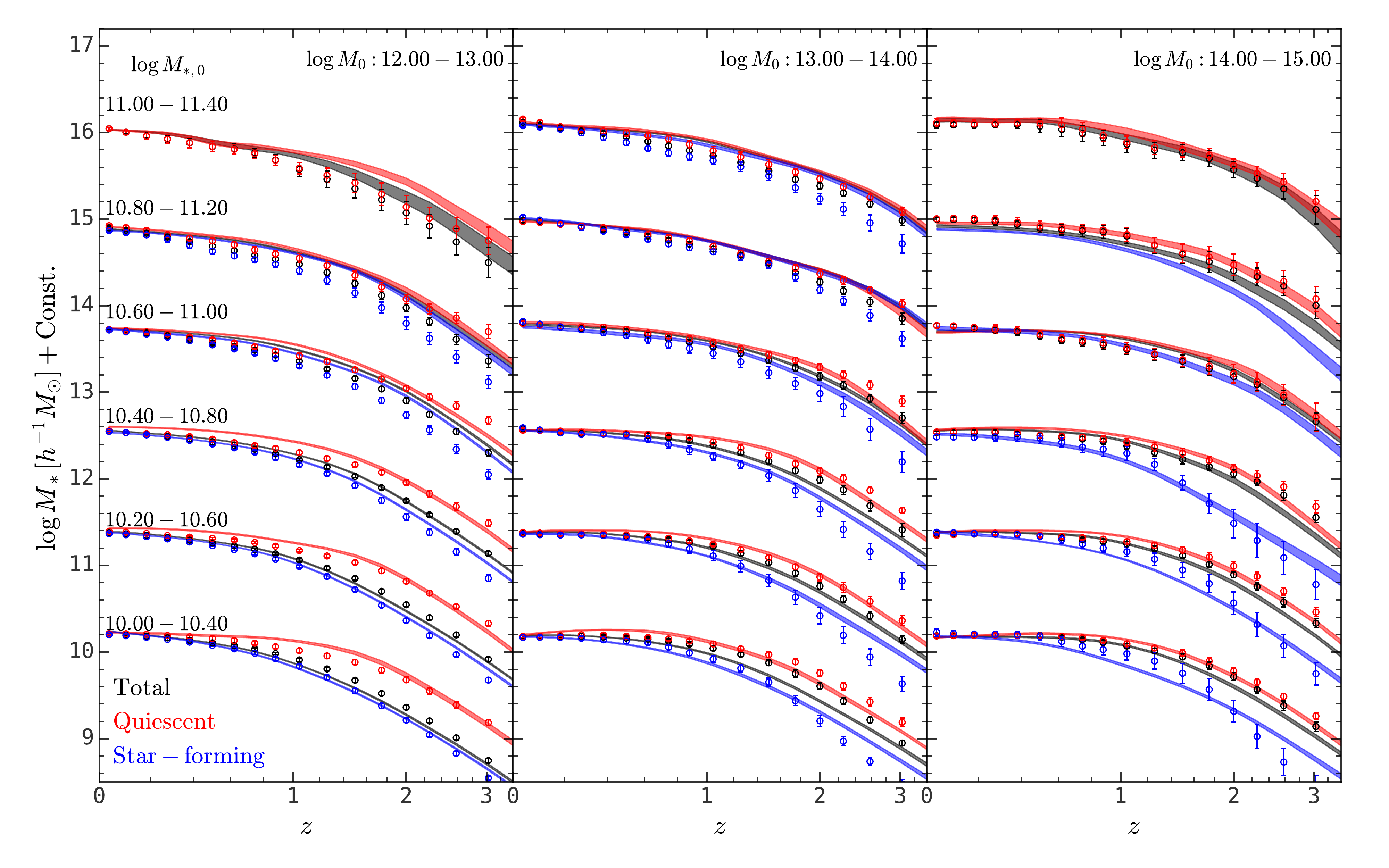}
    \caption{
        Similar to Fig.~\ref{fig:figure/main_progenitor_history_tng100}, just
        for the EAGLE simulation.
    }%
    \label{fig:figure/main_progenitor_history_eagle}
\end{figure*}

To test the robustness of our method, we also apply our method to the EAGLE
simulation \citep{schayeEAGLEProjectSimulating2015a}, which contains a series
of cosmological simulations with state-of-the-art subgrid models with the
GADGET-3 tree-SPH code \citep{springelModellingFeedbackStars2005}. In this
paper, we use the simulation with the identifier of \texttt{Ref-L0100N1504},
which was run with $2\times 1504^3$ particles in a box with a side length
of 100 comoving Mpc. The minimal mass for gas and stellar particles are $1.81\times
10^6M_{\odot}$ and $9.70\times 10^6M_{\odot}$, respectively. The simulation
adopted a flat $\Lambda$CDM cosmology from the {\it Planck} mission
\citep{collaborationPlanck2013Results2014} where $\rm \Omega_m = 0.307$,
$\Omega_{\Lambda} = 0.693$, $\rm \Omega_b = 0.04825$, $\rm\sigma_8=0.8288$, and
$h=0.6777$.

The peak halo mass and $z_{\rm starve}$ are defined in the same way as in the
TNG simulation. The stellar mass we use is the sum of all stellar particles
within 30 physical kpc, and the star formation rate is defined in the same
aperture. We adopt the same $\rho(z)$ as given by equation~(\ref{eq:rho}) for
consistency.

In Fig.~\ref{fig:figure/pnsmf_eagle}, the solid lines show the normalized
stellar mass functions of progenitors extracted from galaxy merger trees in the
EAGLE simulation, and the results derived from our method using summary
statistics of the EAGLE simulation are shown as dashed lines.
Fig.~\ref{fig:figure/main_progenitor_history_eagle} shows the median stellar
mass evolution of main progenitor galaxies in the EAGLE simulation, where the
results extracted from galaxy merger trees are shown by the shaded bands and
the results obtained using our method are shown in symbols. One can see that
our method can properly recover both the progenitor stellar mass distribution
and the evolution of main progenitors.

\bsp	
\label{lastpage}
\end{document}